\documentclass[twocolumn,prb,amsmath,amssymb,showpacs]{revtex4}

\usepackage{graphics}
\usepackage{graphicx}
\usepackage{dcolumn}
\usepackage{bm}
\usepackage{subfigure}

\newcommand*{\Ham}{{\cal{H}}}
\newcommand*{\HamB}{{\cal{H}}_B}
\newcommand*{\HamT}{{\cal{H}}_T}

\newcommand*{\HamL}{{\cal{H}}_{leads}}
\newcommand*{\Heff}{\mbox{\boldmath$H$}}
\newcommand*{\SMM}{\mbox{\boldmath$S$}}
\newcommand*{\svec}{\mbox{\boldmath$\sigma$}}
\newcommand*{\sx}{\sigma_x}
\newcommand*{\sy}{\sigma_y}
\newcommand*{\sz}{\sigma_z}

\newcommand*{\tz}{\hat \tau_3}
\newcommand*{\sv}{\vec \sigma}
\newcommand*{\ex}{\mbox{\boldmath$e$}_x}
\newcommand*{\ey}{\mbox{\boldmath$e$}_y}
\newcommand*{\ez}{\mbox{\boldmath$e$}_z}
\newcommand*{\eS}{\mbox{\boldmath$e$}_S}
\newcommand*{\vecmu}{\mbox{\boldmath$\mu$}}
\newcommand*{\dvec}{\mbox{\boldmath$d$}}
\newcommand*{\gvec}{\mbox{\boldmath$g$}}
\newcommand*{\fvec}{\mbox{\boldmath$f$}}
\newcommand*{\jvec}{\mbox{\boldmath$j$}}
\newcommand*{\jsvec}{\mbox{\boldmath$j$}}

\newcommand*{\vecR}{\mbox{\boldmath$R$}}

\newcommand*{\vheff}{\mbox{\boldmath$h$}_{eff}}
\newcommand*{\pF}{\mbox{\boldmath$p$}_F}
\newcommand*{\vF}{\mbox{\boldmath$v$}_F}
\newcommand*{\torque}{\mbox{\boldmath$\tau$}}

\newcommand*{\Mvec}{\mbox{\boldmath$M$}}

\newcommand*{\Mag}{\mbox{\boldmath$M$}}
\newcommand*{\gammavec}{\mbox{\boldmath$\gamma$}}
\newcommand*{\phivec}{\mbox{\boldmath$\phi$}}

\def\spinup{\uparrow}
\def\spindown{\downarrow}
\def\bvvf{{\mbox{\boldmath$v$}_{F}}}

\def\om{\omega}
\def\be{\begin{equation}}
\def\ee{\end{equation}}
\def\bea{\begin{eqnarray}}
\def\eea{\end{eqnarray}}
\def\bse{\begin{subequations}}
\def\ese{\end{subequations}}
\def\bc{\begin{center}}
\def\ec{\end{center}}

\begin{document}

\title{Non-equilibrium effects in a Josephson junction coupled to a precessing spin}
\author{C. Holmqvist$^1$, S. Teber$^2$ and M. Fogelstr\"om$^1$}
\affiliation{$^1$Department of Microtechnology and Nanoscience - MC2, Chalmers University of Technology,
S-412 96 G\"oteborg, Sweden.\\
$^2$Laboratoire de Physique Th\'eorique et Hautes Energies, Universit\'e Pierre et Marie Curie, 4 place Jussieu, 75005, Paris, France}

\date{\today}

\begin{abstract}
We present a theoretical study of a Josephson junction consisting of two s-wave superconducting leads coupled over a classical spin. 
When an external magnetic field is applied, the classical spin will precess with the Larmor frequency. 
This magnetically active interface results in a time-dependent boundary condition with different tunneling 
amplitudes for spin-up and spin-down quasiparticles and where the precession produces spin-flip scattering processes. 
We show that as a result, the Andreev states develop sidebands and a non-equilibrium population which depend on the 
precession frequency and the angle between the classical spin and the external magnetic field. 
The Andreev states lead to a steady-state Josephson current whose current-phase relation could be 
used for characterizing the precessing spin. 
In addition to the charge transport, a magnetization current is also generated. 
This spin current is time-dependent and its polarization axis rotates with the same precession frequency as the classical spin.
\end{abstract}

\maketitle

\section{Introduction}
\label{sec:intro}

Recently, superconducting-ferromagnetic (SF) hybrid devices have received increased attention due to their potential as spintronics devices. In spintronics, the spin degree of freedom is employed to create new phenomena which could be used to create entirely new devices or be used in combination with conventional charge-based electronics \cite{wolf2001,prinz1998}. Information, e.g., can be stored in the magnetization direction of a small ferromagnet and its state can be read out by measuring a current through a nano-scaled contact determined by the magnetization direction. Nanomagnets such as single molecular magnets or magnetic nanoparticles may be suitable building blocks for such information storage \cite{sanvito2006,bogani2008}.

The interest in single molecular magnets been sparked by their appealingly long relaxation times at low temperatures \cite{christou2000} and experimental breakthroughs in contacting molecules to both superconducting and normal leads has made molecular spintronics a growing field of research. Transport measurements of molecular magnets in normal junctions have been made as a means to characterize the magnetic states \cite{heersche2006,jo2006}. Contacting of C$_{60}$ molecules \cite{winkelmann2009}, metallofullerenes \cite{kasumov2005} and carbon nanotubes \cite{cleuziou2006,jarillo2006} to superconducting leads has also been demonstrated. In addition, single molecular magnets have been suggested for quantum computing applications \cite{leuenberger2001,ardavan2007} due to their long relaxation times.

Currents are not only used to read out the state of a magnet but are also used to control the magnetization direction. Spin-polarized currents carry angular momentum. However, a spin current is not a conserved quantity in a ferromagnet and a spin current oriented in such a way that its direction is perpendicular to the interface plane between the ferromagnetic layer and the leads may lose some of its spin-angular momentum. The angular momentum lost by spin-polarized electrons transported through a ferromagnet is transferred to the ferromagnet. This transfer of angular momentum generates a torque acting on the feromagnet's magnetization direction. This spin-transfer torque mediated by electrical currents was theoretically investigated by Slonczewski \cite{slonczewski1996} and Berger \cite{berger1996} who worked out a description for ferromagnet-normal metal (FN) multilayer structures and showed that spin-transfer torques can lead to precession as well as reversal of the magnetization direction. These theoretical predictions were experimentally verified by Tsoi \cite{tsoi1998} and Myers \cite{myers1999}. Nonequilbrium magnetization dynamics \cite{waintal2001,waintal2002} and spin-transfer torques \cite{zhao2008} in FNF trilayers coupled to superconducting leads have also been studied.

\begin{figure}[b]
\includegraphics[width=0.6\columnwidth,angle=0]{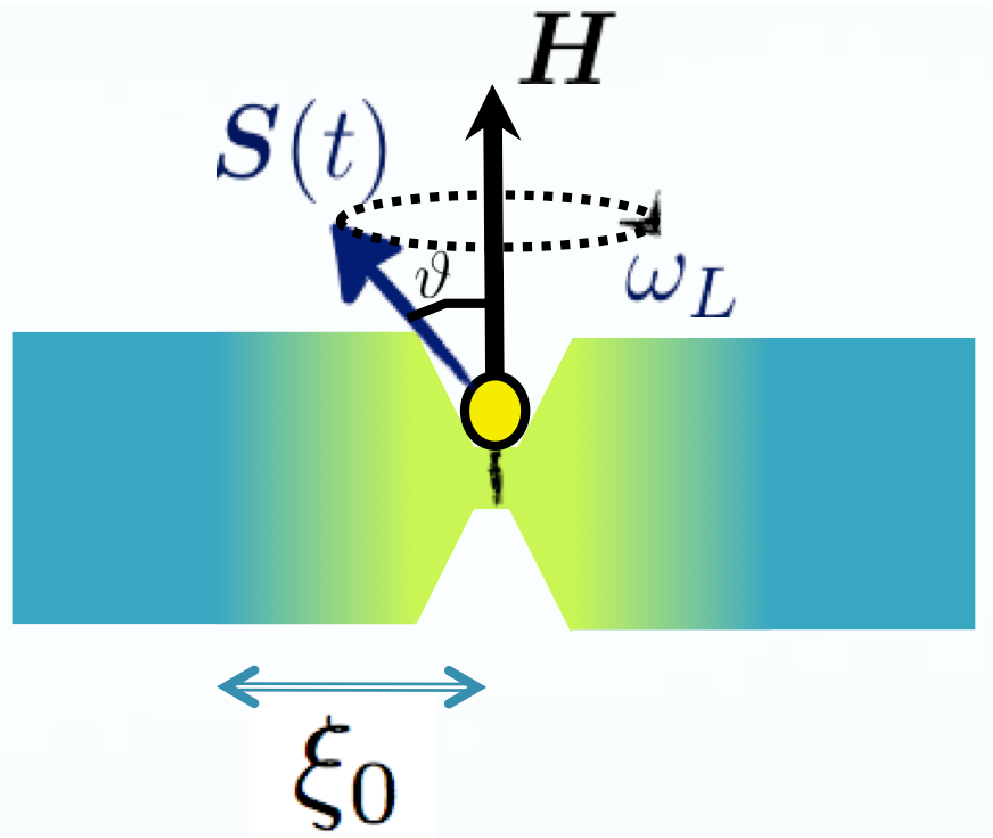}
\caption{(Color online) Two superconducting leads are coupled over a magnetic spin, $\SMM(t)$. The spin precesses with the Larmor frequency $\omega_L$ due to an applied external magnetic field $\Heff$ applied at an angle $\vartheta$. Quasiparticles tunneling between superconductors interact with the precessing spin via an exchange coupling resulting in a dynamical inverse proximity effect existing within the superconducting coherence length $\xi_0$ of the junction interface.}
\label{fig:system}
\vspace*{-0.3truecm}
\end{figure} 

In this paper, we study the coupling between the magnetization dynamics of a nanomagnet or single molecular magnet and Josephson currents through a nano-scaled junction. We consider two superconducting leads coupled over a nanomagnet, consisting of e.g. a molecular magnet or a magnetic nanoparticle as shown in figure \ref{fig:system}. The spins of the magnetic molecule or nanoparticle are assumed to be held parallel to each other resulting in a uniform magnetization which can be represented by a macrospin \cite{stoner1948}. If an external magnetic field is applied, the spin of the nanomagnet starts to precess with the Larmor frequency. This dynamics changes however when it is coupled to conduction electrons in the leads \cite{zhu2003,zhu2004}. 
As our starting point, we take the model by Zhu {\it et al.} \cite{zhu2003,zhu2004} and extend it to include arbitrary tunneling strengths leading to a modified quasiparticle spectrum displaying Andreev levels for energies within the superconducting gap, $\vert \varepsilon \vert < \Delta(T)$ \cite{kulik1969,ishii1970}.
In reference [\onlinecite{teber2010}], we focused on the dc Josephson charge current, while here, we focus on the coupling between the dynamics of the Andreev levels and the transport properties.
The coupling of the two superconducting leads over the precessing spin produces an ac spin Josephson current. The difference between the spin currents on the left and right sides of the interfaces produces a spin-transfer torque, $\torque=\jsvec^{s}_{L}-\jsvec^{s}_{R}$, shifting the precession frequency of the rotating spin.
At finite temperatures, there is also a spin current carried by quasiparticles generating a damping of the magnetization dynamics, the so-called Gilbert damping \cite{gilbert2004,tserkovnyak2002,waintal2003}. A transition of the leads from the normal state into the 
superconducting state reduces the Gilbert damping \cite{bell2008} since the number of 
quasiparticles is suppressed for temperatures $T<T_{\rm c}$ \cite{morten2008}. 
This interplay between the Josephson effect and a
single spin may be used for read-out of quantum-spin states \cite{bulaevskii2004} or for manipulation of
Andreev levels in the junction \cite{michelsen2008}. 

Furthermore, we find that the ac Josephson spin current is a result of superconducting spin-triplet correlations induced by the spin precession.
The appearance of superconducting spin-triplet correlations in SFS junctions has been used to explain the observation of a long-range proximity effect in a number of experiments \cite{keizer2006,sosnin2006,khaire2010,wang2010}. Keizer {\it et al.} observed a supercurrent through a junction consisting of conventional s-wave superconductors coupled over a layer of the half-metallic ferromagnet CrO$_2$ much thicker than the decay length of the superconducting spin-singlet correlations \cite{keizer2006}. Various mechanisms for converting the spin-singlet correlations of the superconducting leads into spin-triplet correlations that may survive within a ferromagnetic layer have been suggested \cite{bergeret2001,eschrig2008,houzet2007}. Bergeret {\it et al.} showed that a local inhomogeneous magnetization direction at the SF interface is sufficient to generate spin-triplet conversion \cite{bergeret2001}. In reference [\onlinecite{eschrig2008}], it was suggested that the spin-singlet to spin-triplet conversion is due to interface regions with misaligned averaged magnetic moments breaking the spin-rotation symmetry of the junction producing spin mixing as well as spin-flip processes. A similar trilayer structure with noncollinear magnetizations resulting in a long-range triplet proximity effect was proposed by Houzet and Buzdin \cite{houzet2007}. Taking into account the importance of interface composition, Khaire {\it et al.} \cite{khaire2010} devised SFS junctions consisting of conventional superconductors and CrO$_2$ in which they had inserted weakly ferromagnetic layers between the superconductors and the half metal to produce interface layers with misaligned magnetization directions. A long-range proximity effect was observed in junctions containing the interface layers, but not in junctions without. Confirmation of Keizer's results were made by Wang {\it et al.} \cite{wang2010} who measured a supercurrent through a crystalline Co nanowire. The Co nanowire was a single crystal, but the contacting procedure was likely to cause defects at the SF interfaces and the inhomogeneous magnetic moments needed to create the spin-triplet correlations. Other experimental verifications of long-range proximity effects includes Holmium (Ho) wires contacted to conventional superconductors \cite{sosnin2006}. Ho has a conical ferromagnetic structure whose magnetization rotates like a helix along the $c$ axis. The appearance of spin-triplet correlations in such junctions and their effect on the long-range proximity effect \cite{linder2009, halasz2009} and spin currents \cite{alidoust2010} have also been studied theoretically. In the present problem, the magnetization direction {\it varies in time} rather than in space giving rise to time-dependent Andreev level dynamics and a dynamical inverse proximity effect in the form of induced time-dependent spin-triplet correlations.
Houzet \cite{houzet08} studied a related problem in which a Josephson junction consisting a ferromagnetic layer with a precessing magnetization placed between two diffusive superconductors was predicted to display a long-range triplet proximity effect.

We formulate the problem of two superconducting leads coupled over a nanomagnet in terms of nonequilibrium Green's functions.
The quasiclassical theory of superconductivity is based on Landau's Fermi liquid theory \cite{landau1957,landau1959} and is applicable to both superconducting \cite{eilenberger1968,larkin1968,eliashberg1971} and superfluid \cite{serene1983} phenomena as well as inhomogeneous superconductors and nonequilibrium situations. The quasiclassical theory gives a macroscopic description where microscopic details are entered as phenomenological parameters \cite{serene1983}. Basically, it is an expansion in a small parameter $k_BT/E_F$, where $E_F$ is the Fermi energy, and is suitable for weakly perturbed superconductors. The perturbations should be weak compared to the Fermi energy, $V\ll E_F$, and of low frequency, $\hbar\omega\ll E_F$.
Interfaces and surfaces in superconducting heterostructures or point contacts, on the other hand, are strong, localized perturbations with strengths comparable to the Fermi surface energy \cite{serene1983}. Within quasiclassical theory, interfaces are handled by formulation of boundary conditions which
usually have been expressed as scattering problems, being able to treat spin-independent 
\cite{zaitsev84,shelankov84,millis88,nagai88,eschrig00,shelankov00} as well as
spin-dependent, or spin-active, interfaces \cite{fogelstrom00,cuevas01,andersson02,barash02,eschrig03,zhao04}.
In many problems, in particular when an explicit time dependence appears,
the T-matrix formulation is more convenient \cite{cuevas96,cuevas01,andersson02}.
This formulation is also well suited for studying interfaces with different numbers of trajectories on either side as is the case for normal metal/half metal interfaces \cite{eschrig03,kopu04}.
The two methods have proved to be equivalent and may be applied both in the limit of clean and in the limit of diffuse superconductors \cite{kopu04}.
In the latter case, the boundary conditions coincides with those of Kuprianov and Lukichev \cite{kuprianov88} and of Nazarov \cite{nazarov99}. For a recent review of quasiclassical theory we refer
the reader to reference [\onlinecite{eschrig09}]. 
In the present problem, the dynamics of the nanomagnet constitutes a time-dependent spin-active boundary condition for the two superconductors which we solve using the T-matrix formulation. First, the transport equations are solved separately to find the classical trajectories for each lead. Then the T-matrix describing the scattering between the leads is used to connect the trajectories across the time-dependent spin-active interface.

We start by outlining the T-matrix formulation applicable to scattering via the precessing magnetic moment in Sec. II. We show that the boundary condition can be solved both in the laboratory frame and a rotating frame. In the latter solution, the explicit time dependence is removed by a transformation to a rotating frame rendering this approach suitable for efficient numerical implementations for studies of transport properties. However, the solution comes at the cost of introducing an energy shift of the chemical potentials for the spin-up and spin-down bands. The laboratory frame approach is, on the other hand, suitable for studying modifications to the superconducting state although the explicit time dependence increases the complexity of the solution. In Sec. III A, we review the results for the Josephson charge current in reference [\onlinecite{teber2010}] in terms of the laboratory frame description. The spin currents are described in Sec. III B, which is followed in Sec. III C by the induced time-dependent spin-triplet correlations and Andreev-level dynamics giving rise to the spin currents. In Sec. III D, we discuss the back-action of the scattering processes on the magnetization dynamics while the magnetization induced in the leads is discussed in section III E. In Sec. IV, we conclude with a summary of our results.

\section{Model}
\label{sec:model}

We consider two superconductors forming a Josephson junction over a nanomagnet. The nanomagnet may either be magnetic 
nanoparticle or a single-molecule magnet and we will assume that contact between the leads and the nanomagnet is made up of 
a few single quantum channels. The magnetization of the nanomagnet is put in precession and the resulting contact will constitute a time-dependent 
spin-active interface (see figure \ref{fig:system}). The nanomagnet together with the two superconducting leads are described by 
the total Hamiltonian \cite{zhu2003,zhu2004}
\begin{equation}
\Ham=\HamL+\HamB+\HamT.
\end{equation}
The left (L) and right (R) leads are s-wave superconductors described by the BCS Hamiltonian
\begin{equation}
\HamL=\sum_{k,\sigma \atop \alpha=L,R} \xi_k c^\dagger_{\alpha,k,\sigma} c_{\alpha,k,\sigma}
+\sum_{k \atop \alpha=L,R} \lbrack \Delta_\alpha c^\dagger_{\alpha,k,\uparrow}c^\dagger_{\alpha,-k,\downarrow}+h.c.\rbrack
\end{equation}
where the dispersion, $\xi_k=\hbar^2k^2/2m-\mu$, and the chemical potential, $\mu$, are assumed to be the same for both leads. The order parameter of the leads is assumed temperature dependent, $\Delta_\alpha=\Delta(T){\rm e}^{\pm i \varphi /2}$. Here $\varphi$ is
the relative superconducting phase difference over the junction which we treat as a static variable that is tunable.
The nanomagnet is subjected to an external magnetic field modeled as an effective field $\Heff$ acting on the nanomagnet's magnetic moment, $\vecmu$. Included in this effective field are also any r.f. fields to maintain precession, crystal anisotropy fields and demagnetization effects. The magnetic moment of the nanomagnet is viewed as a single spin, or macrospin, which we will treat as a classical entity. This macrospin is related to the magnetic moment by $\vecmu=-\gamma\SMM$ where $\gamma$ is the gyromagnetic ratio. The spin and the effective magnetic field couple via a Zeeman term,
\begin{equation}
\HamB=-\gamma \SMM\cdot\Heff.
\end{equation}
If the effective field is applied at an angle, $\vartheta$, relative to the spin, a torque is produced that brings the classical spin into precession around the direction of the effective field. This precession generated by the tilt angle occurs with the Larmor frequency $\omega_L=\gamma H$, where the magnitude of the external field is $H=\vert \Heff \vert$. Here, we take the direction of the effective field to be along the $\ez$ axis and the angle $\vartheta$ is defined as $\Heff\cdot\SMM(t)= H S \cos \vartheta$.
The Larmor precession is captured by the equation of motion
\begin{equation}
\frac{d\SMM}{dt}=-\gamma \SMM \times \Heff
\end{equation}
where the right-hand side is the torque produced by the effective field. 
The magnitude of the spin is constant, $S=\vert \SMM (t) \vert={\rm const}$. Consequently, the path traced out by the spin, $\SMM(t)=S \eS (t)$, has a time dependence which lies totally in the direction of the spin, $\eS (t)$. In the absence of any other torques than the effective field, the direction of the precessing spin may be written as
\begin{equation}
\eS (t)= \big(\cos (\omega_L t)  \sin \vartheta\, \ex+\sin (\omega_L t)  \sin \vartheta\,\ey+\cos \vartheta\, \ez \big).
\end{equation}
Tunneling quasiparticles have the possibility to tunnel directly between the leads with hopping amplitude $V_0$ or interact with the precessing spin via an exchange coupling of strength $V_S$. These processes are described by the tunneling Hamiltonian
\begin{equation}
\HamT= \sum_{k \sigma;k^\prime\sigma^\prime} c^\dagger_{L,k\sigma} V_{k\sigma;k^\prime\sigma^\prime} c_{R,k^\prime\sigma^\prime} +
c^\dagger_{R,k^\prime\sigma^\prime} V^\dagger_{k\sigma;k^\prime\sigma^\prime} c_{L,k\sigma},
\label{Htunnel}
\end{equation}
where $V_{k\sigma;k^\prime \sigma^{\prime}}=(V_0 \delta_{\sigma \sigma^{\prime}}+V_S(\SMM (t) \cdot \svec)_{\sigma \sigma^{\prime}})\delta(k-k^{\prime})$ and $\svec=(\sx,\sy,\sz)$ with $\sigma_i$ being the Pauli matrices. The spin-dependent hopping amplitude is
\begin{equation} 
V_S\SMM(t)\cdot \svec = V_SS \big (\cos \vartheta\, \sz+ \sin \vartheta\, {\rm e}^{-i \omega_L t \sz}\sx\big)
\end{equation}
where the first term is a spin-conserving part with different hopping amplitudes for spin-up and spin-down quasiparticles. The second term is time-dependent and describes processes where quasiparticles flip their spins while exchanging energy $\omega_L$ with the rotating spin.
The junction's transport properties as well as the modifications to the superconducting states of the leads depend on the hopping amplitudes $V_0,V_S$ as well as the the superconducting phase difference, $\varphi$, the precession of the spin, $\SMM(t)$, and the tilt angle, $\vartheta$, between the effective field and the precessing spin.

\subsection{Approach}
\label{sec:approach}

We formulate the problem using nonequilibrium Green's functions in the quasiclassical approximation following references [\onlinecite{cuevas01,kopu04}]. The tunneling Hamiltonian (\ref{Htunnel}) provides a time-dependent and spin-active boundary condition for the quasiclassical Green's function and is solved by a T-matrix equation \cite{caroli71,buchholtz79,martinrodero94}.
A quasiclassical Green's function is a propagator describing quasiparticles moving along classical trajectories defined by the
Fermi velocity $\bvvf=\bvvf(\pF)$ at a given quasiparticle momentum on the Fermi surface, $\pF$. 
The information of a quasiclassical Green's function is contained in the the object
\begin{equation}
\check g(\pF,\vecR;\varepsilon,t)=\left(\begin{array}{cc} \hat g^R(\pF,\vecR;\varepsilon,t) &\hat g^K(\pF,\vecR;\varepsilon,t) \\
0 &\hat g^A(\pF,\vecR;\varepsilon,t) \end{array}\right)
\end{equation}
where $\vecR$ is the spatial coordinate, $\varepsilon$ is the quasiparticle energy relative to the chemical potential $\mu$ and $t$ is time. The $\check g(\pF,\vecR;\varepsilon,t)$ propagator is a $8\times8$ matrix in the combined Keldysh-Nambu-spin space. The "check" denotes a $2\times2$ matrix in Keldysh space where the components are retarded ($R$), advanced ($A$) and Keldysh ($K$) Green's functions while the "hat" indicates a $2\times2$ matrix in Nambu, or particle-hole, space which is further parameterized using the Pauli spin matrices ($\sx,\sy,\sz$) (see reference [\onlinecite{serene1983}] for details). The matrix components in the combined Nambu-spin space are conveniently divided into spin scalar ($s$) and spin vector ($t$) parts,
\begin{equation}\label{scalar and triplet green's function}
\hat{g}^{X}=\left(\begin{array}{cc}g^X_s+ \gvec^X_t \cdot \svec & (f^X_s+\fvec^X_t\cdot\svec)i\sy \\ i\sy ({\tilde f}^X_s + {\tilde 
\fvec}^X_t \cdot \svec)& {\tilde g}^X_s - \sigma_y ({\tilde \gvec}^X_t\cdot \svec )\sigma_y  \end{array}\right)
\end{equation}
$(X=R,A,K)$ where $f^X_s$ and $\fvec^X_t$ are the spin-singlet and spin-triplet components of the anomalous Green's functions (similarly for $g^X_s$ and $\gvec^X_t$). The Green's function $\check g(\pF,\vecR;\varepsilon,t)$ obeys a Boltzman-like transport equation \cite{eilenberger1968,larkin1968,eliashberg1971}
\begin{equation}
i \hbar \vF\!\cdot\!\nabla \check g+\lbrack \varepsilon \tz \check 1 -\check{\Ham}, \check g\rbrack_\circ=0,
\label{transpeq}
\end{equation}
where $\hat \tau_i$ are Pauli matrices in Nambu space and $\check \Ham$ includes self-energies like the superconducting order parameter, $\check{\Delta}=\hat{\Delta}\check{1}$, and impurity contributions, $\check \Sigma_{imp}$, as well as any external fields, $\check{h}_{ext}$. The "$\circ$" product represents a convolution over common time arguments combined with a matrix muliplication \cite{serene1983,eschrig09}. The transport equation is complemented by the Eilenberger normalization condition,
\begin{equation}
\check g \circ \check g=-\pi^2 \check 1,
\label{normcond}
\end{equation}
and a set of self-consistency equations such as the one for the superconducting order parameter,
\begin{equation}
\hat{\Delta}(\vecR,t)=\lambda\int_{-\varepsilon_c}^{\varepsilon_c}\frac{d\varepsilon}{4\pi i}\langle \hat{f}^K(\pF,\vecR;\varepsilon,t) \rangle_{\pF},
\label{BCSgapequation}
\end{equation}
where $\langle \cdot \rangle_{\pF}$ is an average over the Fermi surface, $\lambda$ is the pairing-interaction strength and $\varepsilon_c$ is the cut-off energy which may be eliminated by making use of the critical temperature, $T_c$.

The transport equation (\ref{transpeq}) is solved separately for each lead, treating the interface as an impenetrable surface where quasiparticles are perfectly reflected. This hard wall boundary condition leads to a solution, $\check{g}^0_{\alpha}$, for each semi-infinite lead, $\alpha=L,R$.
The propagators are then connected across the interface by the tunneling Hamiltonian, $\HamT$, whose effects can be incorporated via a quasiclassical t-matrix equation as
\begin{equation}
\check t_\alpha(t,t^\prime)=\check \Gamma_\alpha(t,t^\prime) 
+\lbrack \check \Gamma_\alpha\!\circ\!\check g^{0}_\alpha \!\circ\!  \check t_\alpha\rbrack (t,t^\prime).
\label{boundaryTmatrix}
\end{equation}
The hopping elements of the tunneling Hamiltonian enter the t-matrix equation via a matrix $\check \Gamma_{L}(t,t^\prime)$ defined as
\begin{equation}
\check \Gamma_{L}(t,t^\prime)=\lbrack \check v \!\circ\! \check g^{0}_{R}\!\circ\!\check v\rbrack (t,t^\prime)
\label{gammaM}
\end{equation}
for the left side of the interface and the right-side matrix $\check \Gamma_{R}$ is obtained by interchanging $L$ and $R$. The time dependence in the current problem enters through the hopping element $\check{v}(t)$, which in particle-hole space has the form
\begin{equation}
\hat v(t)=\left(\begin{array}{cc} 
v_0+v_S \eS(t)\!\cdot\!\svec &0\\ 
0 &v_0-v_S \sy(\eS(t)\!\cdot\!\svec)\sy
\end{array}\right)
\label{hopping}
\end{equation}
and $\check v(t) =\hat v(t) \check 1$ in Keldysh space. Here, the hopping elements $V_0,V_S$ in equation (\ref{Htunnel}) have been replaced with their Fermi-surface limits, $v_0=\pi N_F V_0$ and $v_S=\pi N_F S V_S$ with $N_F$ being the normal density of states at $E_F$. Note that for the junction studied, the hopping elements have the symmetries $\hat v_{LR}(t)=\hat v_{LR}^\dagger(t)=\hat v_{RL}(t)\equiv\hat v(t)$.
The t matrices (\ref{boundaryTmatrix}) are used to calculate the full quasiclassical propagators which depending on if their trajectories lead up to or away from the interface are divided into "incoming" ($\check g^{i}$) and "outgoing" ($\check g^{o}$) propagators, given by
\begin{equation}
\check g^{i,o}_\alpha(t,t^\prime)=\check g^{0}_\alpha(t,t^\prime)+
\lbrack (\check g^{0}_\alpha\pm i \pi \check 1)\!\circ\! t_\alpha \!\circ\!  (\check g^{0}_\alpha\mp i \pi \check 1)\rbrack(t,t^\prime)
\label{boundary}
\end{equation}
where $\alpha=L,R$ and the upper and lower signs, $\pm$ and $\mp$, refer to the incoming and outgoing propagators, respectively.
 
The self-energy fields, such as the order parameter, depend on the full propagators and should in principle be calculated self-consistenly taking into account the interface scattering. However, we assume in this study that the area of the point contact, ${\cal {A}}$, is small compared to $\pi\xi_0^2$ where $\xi_0=\hbar \vert \vF \vert /2\pi T_c$ is the superconducting coherence length. As a result, the superconducting state does not change considerably and the order parameter, $\check\Delta$, and other possible self-energies in the leads do not have to be recalculated \cite{serene1983}. We will also assume that the superconducting phase changes abruptly over the contact.

The use of equations (\ref{transpeq},\ref{normcond}) together with the boundary condition (\ref{boundary}) allows for calculation of the transport properties the junction.
The charge and spin currents are given by an average over the Fermi-surface momentum directions of the full propagators. Here, this average amounts to a difference between incoming and outgoing propagators and the charge, $j^{c}$, and spin, $\jvec^{s}$, currents evaluated in lead $\alpha$ for a single conduction channel are
\begin{equation}
j^{c}_\alpha(t) = \frac{e}{2\hbar} \int \frac{d \varepsilon}{8 \pi i} \mbox{Tr} [ \hat{\tau}_3 
(\hat{g}^{i,<}_{\alpha}(\varepsilon,t)  - \hat{g}^{o,<}_{\alpha}(\varepsilon,t)) ]  
\label{chargecurr}
\end{equation}
and
\begin{equation}
\jvec^{s}_\alpha(t) =  \frac{1}{4}  \int \frac{d \varepsilon}{8 \pi i}  \mbox{Tr} [ \hat{\tau}_3 \hat{\svec} 
( \hat{g}^{i,<}_{\alpha}(\varepsilon,t) - \hat{g}^{o,<}_{\alpha}(\varepsilon,t)) ],
\label{spincurr}
\end{equation}
where $\hat{\svec}={\rm diag}(\svec,-\sy\svec\sy)$. The Green's functions in the above expressions are the lesser propagators defined as $\hat{g}^<=\frac{1}{2}(\hat{g}^{K}-\hat{g}^{R}+\hat{g}^{A})$.

\subsection{Solving the time-dependent boundary condition}
The time-dependent boundary conditions can be solved in two different ways. In the first procedure, the boundary conditions are solved in the laboratory frame in which the time dependence is preserved and is manifested as frequency shifts in a difference equation. The treatment is similar to that of dc-voltage biased SIS junctions \cite{arnold87,bratus95,averin95,cuevas96,cuevas01,andersson02,zhao2008}. The second approach involves removing the explicit time dependence of $\hat{v}(t)$ by a transformation to a rotating frame (see reference [\onlinecite{teber2010}] for details). This procedure is numerically more efficient but the transformation, however, introduces an exchange field shifting the chemical potentials of the spin-up and spin-down bands in the leads making the first approach more suitable for studying changes in the superconducting state in vicinity of the junction due to quasiparticle tunneling via the precessing spin. Below we describe both within a quasiclassical framework.

\subsubsection{Laboratory frame}

To solve the boundary condition (\ref{boundary}) dependent on the matrix (\ref{gammaM}) in the laboratory frame, it is more convenient to Fourier transform the t-matrix equation from the time domain to energy space where it becomes an algebraic equation,
\begin{equation}
\check t_\alpha(\varepsilon,\varepsilon^{\prime})=\check \Gamma_\alpha(\varepsilon,\varepsilon^{\prime})
+\sum_{\varepsilon^{''}}\check \Gamma_\alpha(\varepsilon,\varepsilon^{''}) 
\check {g}^0_{\alpha} (\varepsilon^{''})\check t_\alpha(\varepsilon^{''},\varepsilon^{\prime}).
\label{boundaryE}
\end{equation}
The propagators $\check g^0_{\alpha}(\varepsilon,\varepsilon^\prime)=\check g^0_{\alpha}(\varepsilon)\delta(\varepsilon-\varepsilon^\prime)$ have the following Nambu-spin structure,
\begin{eqnarray}
&\hat g^{0,R}_{\alpha}(\varepsilon)&=\left(\begin{array}{rl}g^R(\varepsilon)& f^R(\varepsilon)i\sy \\ i\sy {\tilde f}^R(\varepsilon) & {\tilde g}^R(\varepsilon)\end{array}\right)\nonumber\\
&&=-\frac{\pi}{\Omega^R}\left(\begin{array}{cc} \varepsilon^R & \Delta(T,\pm\varphi) i\sy \\ 
i\sy \Delta^*(T,\pm\varphi) &-\varepsilon^R\end{array}\right)\nonumber\\
&{\rm where}& \Omega^R=\sqrt{\vert\Delta(T)\vert^2-(\varepsilon^R)^2},\,\,\varepsilon^R=\varepsilon+i 0^+,
\nonumber\\
&\hat g^{0,A}_{\alpha}(\varepsilon)&=\tz [ \hat g^{0,R}_{\alpha}(\varepsilon) ]^\dagger \tz ,\label{gleads}\\
\nonumber &{\rm and}& \\ 
&\hat g^{0,K}_{\alpha}(\varepsilon)&=(\hat g^{0,R}_{\alpha}(\varepsilon)-\hat g^{0,A}_{\alpha}(\varepsilon))\tanh(\varepsilon/2T).
\nonumber
\end{eqnarray}
The gap, $\Delta(T,\pm\varphi)=\Delta(T){\rm e}^{\pm i \varphi/2}$, is both temperature and phase dependent and the "$+$"("$-$") sign of the phase dependence refers to lead $R$($L$). Furthermore, $\tanh(\varepsilon/2T)$ is the quasiparticle occupation function setting the two superconducting leads in thermal equilibrium with each other.
The t matrices in energy space are a sum of t matrices whose energies differ by $\omega_L$ and satisfy the relation
\begin{equation}
\check{t}_\alpha (\varepsilon, \varepsilon') = \sum_n \check{t}_\alpha (\varepsilon, \varepsilon+n \omega_L) \delta(\varepsilon - \varepsilon' +n \omega_L),
\end{equation}
which is equivalent to a time dependence of
\begin{equation}
\check{t}_\alpha (t, t') = \sum_n e^{-in \omega_L t} \int \frac{d \varepsilon}{2 \pi} e^{-i \varepsilon (t-t')}  \check{t}_\alpha (\varepsilon+n \omega_L, \varepsilon).
\end{equation}
The Ansatz above and the assumption that the leads are in equilibrium so that their respective Green's functions may be written as $\check g^0_\alpha (t, t') =\check g^0_\alpha (t-t')$ lets one evaluate the coefficient matrices $\check \Gamma_\alpha(\varepsilon,\varepsilon^\prime)$ and  $\check \Gamma_\alpha(\varepsilon,\varepsilon^{'}) 
\check {g}^0_{\alpha} (\varepsilon^{'})$ in equation (\ref{boundaryE}) and subsequently solve the resulting difference equation in terms of $\check{t}_\alpha (\varepsilon+n \omega_L, \varepsilon)$ \cite{cuevas96,cuevas01,andersson02}. This is a quite general procedure capable of handling diverse forms of self-energy fields ${\check{\cal {H}}}$. However, the matrix coefficients in equation (\ref{boundaryE}) must be evaluated for each particular kind of junction and lead state. The properties of these matrices then determine the specific t-matrix difference equation and the solution strategy.

For the present calculation certain simplifications can be made due to the spin independence of $\check {g}^0_{\alpha}(\varepsilon)$ given by equation (\ref{gleads}) and the form of the hopping element in equation (\ref{hopping}): the Keldysh-Nambu-spin matrices can be factorized in spin space into generalized diagonal matrices $\check{X}^d$, spin-raising matrices
$\check{X}^\uparrow$, and spin-lowering matrices $\check{X}^\downarrow$. These matrices are still Keldysh-Nambu matrices but they have the algebraic properties of spin matrices such as $\check X^\uparrow\circ\check Y^\uparrow=\check X^\downarrow\circ\check Y^\downarrow=0$,
$\check X^{\downarrow,\uparrow}\circ\check Y^{\uparrow,\downarrow}\propto\check Z^{d}$, and 
$\check X^d\circ\check Y^{\uparrow,\downarrow}\propto\check Z^{\uparrow,\downarrow}$.
A matrix factorized in this form may be shown to have the time dependence
\begin{eqnarray}\label{Xmatrix}
\check X(t,t')&=&\int \frac{d\varepsilon}{2 \pi}{\rm e}^{-i\varepsilon(t-t')}\bigg\lbrack 
\check X^d(\varepsilon,\omega_L)+\\
&+&{\rm e}^{-i \omega_L t}\check X^\uparrow(\varepsilon,\omega_L)+
{\rm e}^{ i \omega_L t}  \check X^\downarrow(\varepsilon,\omega_L)\bigg\rbrack . \nonumber
\end{eqnarray}
Using the spin algebra, the t-matrix equation in energy space (equation (\ref{boundaryE})) may be written as
\begin{equation}\left(\begin{array}{ccc}
\check{1} - \check{A}^d_{1}& - \check{B}^\uparrow_{1} & \check{0} \\
- \check{B}^\downarrow_{0} & \check{1} - \check{A}^d_{0} & - \check{B}^\uparrow_{0}  \\
\check{0} & - \check{B}^\downarrow_{-1} &\check{1} - \check{A}^d_{-1} 
\end{array}\right)
\left(\begin{array}{c}
\check{t}^\uparrow \\ \check{t}^d\\ \check{t}^\downarrow \end{array}\right)
=\left(\begin{array}{c} 
\check{\Gamma}^\uparrow \\ \check{\Gamma}^d \\ \check{\Gamma}^\downarrow
\end{array}\right).
\label{ttmatrix}
\end{equation}

The coefficient matrices in equation (\ref{ttmatrix}) are functions of energy $\varepsilon$ and precession frequency $\omega_L$ and
are straight forward to evaluate.  
Contrary to e.g. the case of a finite dc voltage, where the t-matrix equation is a difference equation solved
by recursive methods, we have a matrix equation for $\check t$ which can be solved by simple (numerical) inversion.
Factorizing the propagators in equation (\ref{boundary}) according to their spin structure results in
\begin{eqnarray}
&\!\!\!\!\!\!\check g^{d,i,o}_{\alpha}(\varepsilon)&\!\!=\!\check g^{0}_{\alpha}(\varepsilon)+\check M^{d}_{\alpha,\pm}(\varepsilon)
\check t^{d}_{\alpha}(\varepsilon,\omega_L)\check M^{d}_{\alpha,\mp}(\varepsilon)\label{gdiag}\\
&\!\!\!\!\!\!\check g^{\uparrow,i,o}_{\alpha}(\varepsilon,t)&\!\!=\!{\rm e}^{-i\omega_L t} \check M^{d}_{\alpha,\pm}(\varepsilon\!+\!\omega_L)
\check t^{\uparrow}_{\alpha}(\varepsilon,\omega_L)\check M^{d}_{\alpha,\mp}(\varepsilon)\label{gup}\\
&\!\!\!\!\!\!\check g^{\downarrow,i,o}_{\alpha}(\varepsilon,t)&\!\!=\!{\rm e}^{+i\omega_L t} \check M^{d}_{\alpha,\pm}(\varepsilon\!-\!\omega_L)
\check t^{\downarrow}_{\alpha}(\varepsilon,\omega_L)\check M^{d}_{\alpha,\mp}(\varepsilon)\label{gdown}
\end{eqnarray}
with $\check M^{d}_{\alpha,\pm}(\varepsilon)=\check g^{0}_{\alpha}(\varepsilon)\pm i\pi \check 1$.

\subsubsection{Rotating frame}

The unitary transformation matrix for removing the time dependence of $\hat{v}(t)$ is
\begin{equation}
\hat {\cal {U}}(t)=\left(\begin{array}{cc}{\rm e}^{-i \frac{\omega_L}{2}t \sz} &0 \\0& {\rm e}^{i \frac{\omega_L}{2}t \sz} \end{array}\right)
\label{utrans}
\end{equation}
resulting in a transformation to a rotating frame of reference with respect to the precessing spin $\SMM(t)$ in which the hopping element is
\begin{equation} 
\hat v=\hat {\cal {U}}^\dagger(t) \hat v(t) \hat {\cal {U}}(t)=\left(\begin{array}{cc}
v_0+v_S \eS\!\cdot\!\sv&0\\
0&v_0-v_S \sy\eS\!\cdot\!\sv\sy
\end{array}\right).
\end{equation}
The direction $\eS$ of the precessing spin is now static in this rotating frame, $\eS=\cos \vartheta \ez+\sin \vartheta \ex$, but the hopping element $\hat{v}$ is still spin active with different hopping amplitudes for spin-up and spin-down quasiparticles, $v_0\pm v_S\cos\vartheta$. The hopping element also contains a spin-flip term $v_S\sin\vartheta$ scattering between the two spin bands.

Next, we apply the unitary transformation to the propagator $\check g_\alpha(t,t^\prime)$ and obtain
\begin{eqnarray}
\check {\tilde g}_\alpha(t,t^\prime)&=&\hat {\cal {U}}^\dagger(t)\, \check g_\alpha(t,t^\prime)\, \hat {\cal {U}}(t^\prime)\\
&=&\hat {\cal {U}}^\dagger(t) \int \frac{d\varepsilon}{2\pi}\frac{d\varepsilon^\prime}{2\pi}
{\rm e}^{-i( \varepsilon t-\varepsilon^\prime t^\prime)}
\check g_\alpha(\varepsilon,\varepsilon^\prime)\, \hat {\cal {U}}(t^\prime)\nonumber.
\label{transg}
\end{eqnarray}
The transformation of the propagators into the rotating frame introduces spin-dependent energy shifts, $\varepsilon \rightarrow \varepsilon \pm\omega_L/2$, displayed in the transformed propagators, e.g. as
\begin{eqnarray}
&\,&\hat {\tilde g}^{0,R}_{\alpha}\!(\varepsilon) \\ \nonumber
&=&\left(\begin{array}{cccc}
g^R_{\alpha}\!(\varepsilon\!+\!\frac{\omega_L}{2})\!&0&0& f^R_{\alpha}\!(\varepsilon\!+\!\frac{\omega_L}{2})\! \\
0&g^R_{\alpha}\!(\varepsilon\!-\!\frac{\omega_L}{2})\!& -f^R_{\alpha}\!(\varepsilon\!-\!\frac{\omega_L}{2})\!&0Ê\\
0& {\tilde f}^R_{\alpha}\!(\varepsilon\!-\!\frac{\omega_L}{2})\!&{\tilde g}^R_{\alpha}\!(\varepsilon\!-\!\frac{\omega_L}{2})\!&0Ê\\
 -{\tilde f}^R_{\alpha}\!(\varepsilon\!+\!\frac{\omega_L}{2})\!&0&0&{\tilde g}^R_{\alpha}\!(\varepsilon\!+\!\frac{\omega_L}{2}\!)
\end{array}\right)
\end{eqnarray}
for the retarded Green's function. The advanced and Keldysh propagators are similarly transformed.

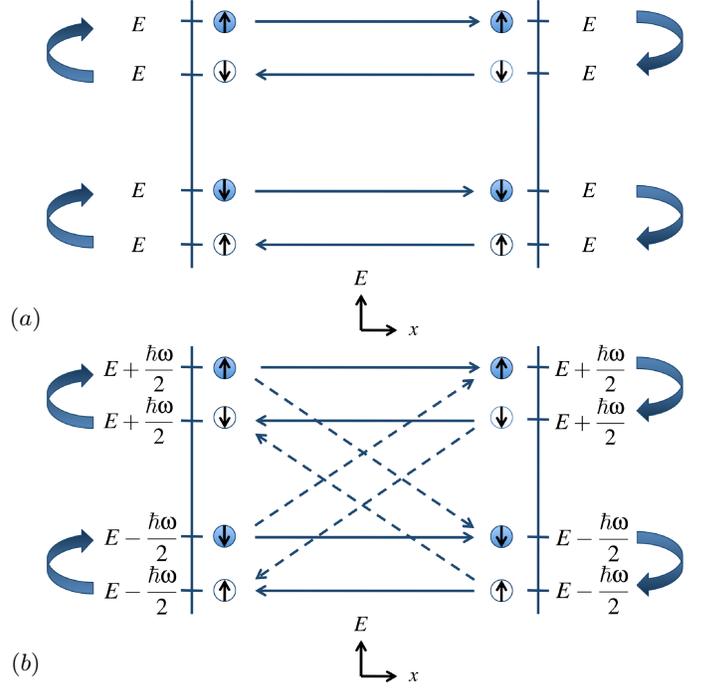
\begin{figure}[t]
  \begin{tabular}{cc}
  \begin{picture}(10,0)
  \put(5,5){\makebox(0,3){$(a)$}}
\end{picture}
  \scalebox{0.115}{\includegraphics{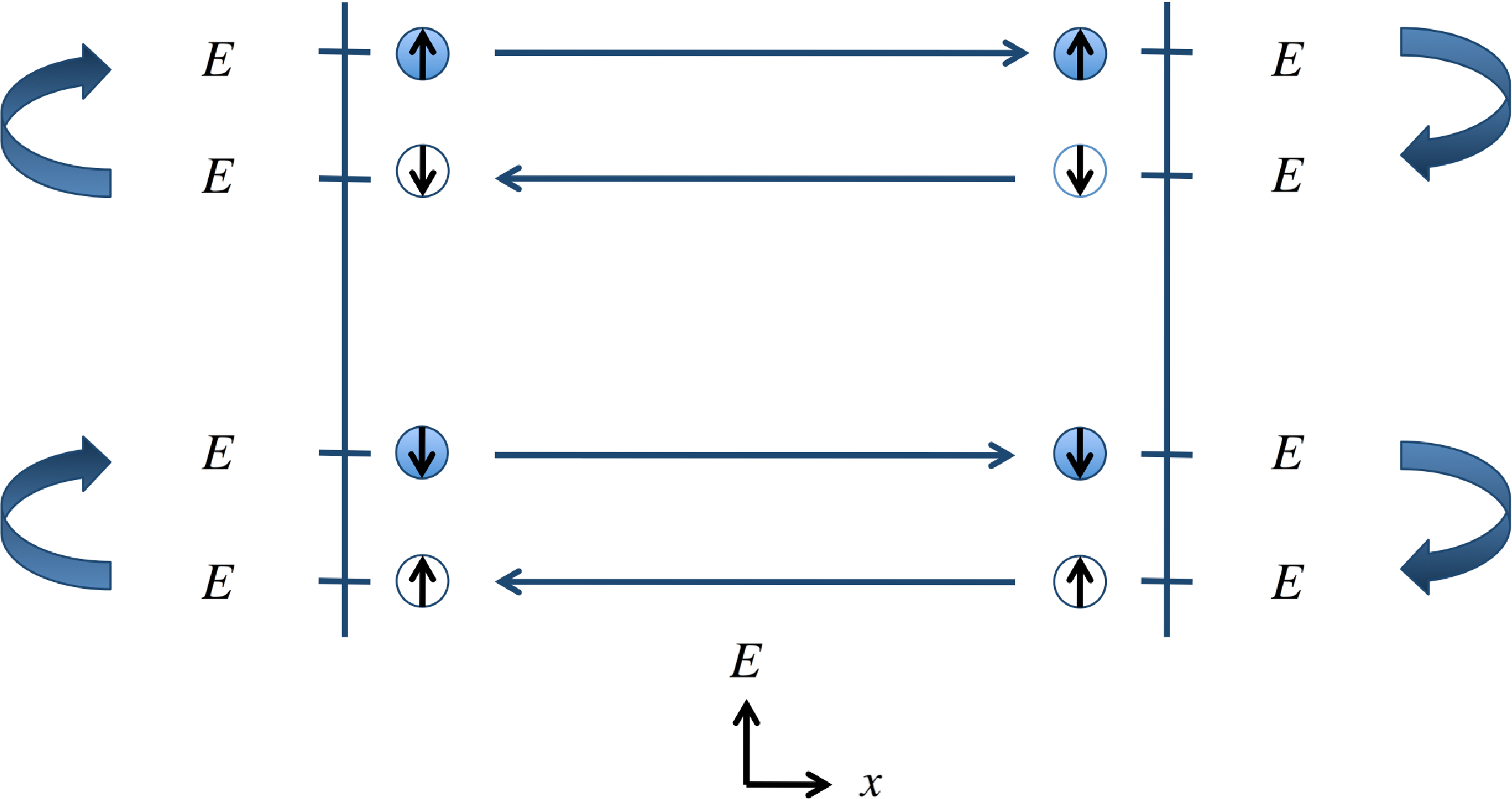}}  
 \\
   \begin{picture}(10,0)
  \put(5,5){\makebox(0,3){$(b)$}}
\end{picture}
 \scalebox{0.115}{\includegraphics{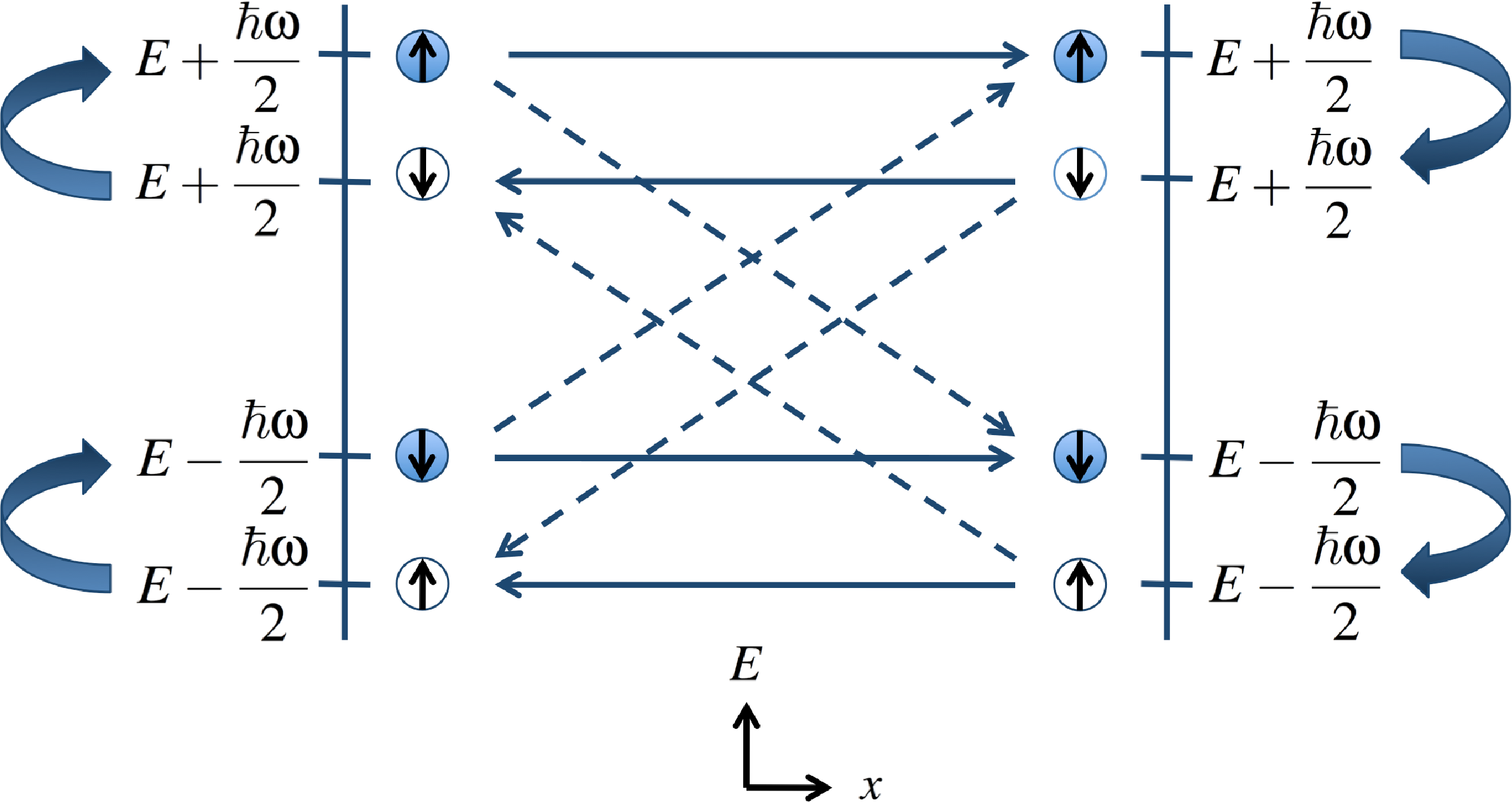}}  
  \end{tabular}
\caption{\label{Scattering processes}
(Color online) (a) Schematic illustration of Andreev scattering in a Josephson junction. Electron-like quasiparticles (blue) impinging on the right superconductor are Andreev retroreflected (thick arrows) into hole-like quasiparticles (white) at the superconductor surface. The hole-like quasiparticles are transmitted (thin arrows) to the opposite (left) superconductor, where they are Andreev retroreflected into electron-like quasiparticles. This Andreev scattering process is phase coherent and results in the formation of quasiparticle states within the superconducting gap, Andreev levels, when the quasiparticles interfere constructively along close loops.
(b) Scattering processes in a Josephson junction coupled coupled over a classical spin precessing with frequency $\omega_L$. A tunneling quasiparticle has the possibility to tunnel across the junction with preserved spin and energy, or having its spin flipped by interaction with the classical spin while gaining or losing energy $\omega_L$. Andreev reflection occurs at the junction interfaces.}
\end{figure}

Applying ${\cal {U}}(t)$ to the transport equation, equation (\ref{transpeq}), similarly leads to a shift of the energies of the spin-up and spin-down quasiparticles. The resulting transport equation is
\begin{equation}
i \hbar \vF\!\cdot\!\nabla \check {\tilde g}_\alpha+\lbrack (\varepsilon \tz +\vheff\!\cdot\!\sz )\check 1 -{\check {\tilde {\cal {H}}}}_\alpha, \check {\tilde g}_\alpha\rbrack_\circ=0,
\label{transpeq, rot frame}
\end{equation}
where the energy shifts are captured by the appearance of an effective magnetic field in the leads, $\vheff=\frac{\omega_L}{2} \ez$. Equation (\ref{transpeq, rot frame}) is time independent as long as
the self-energy fields in ${\check{\cal {H}}}_{\alpha}$ do not contain terms which are off-diagonal in spin space corresponding to spin-flip scattering or equal-spin paring order parameters. For ballistic, or clean, s-wave superconducting leads, ${\check{\cal {H}}}_{\alpha}=\check{\Delta(T)}_{\alpha}$ and equation (\ref{transpeq, rot frame}) is time independent. The propagators $\check {\tilde g}_\alpha(t,t^\prime)$ obey equation (\ref{transpeq, rot frame}) for a specular-scattering boundary condition \cite{kopu04,cuevas01}. Introduction of the tunneling processes lead to a boundary condition problem which is still time independent in this rotating frame and the "$\circ$" product in equation (\ref{boundary}) reduces to a matrix multiplication in energy space.

The precessing spin introduces spin mixing of the two spin bands in such a way that spin-up(-down) quasiparticles are scattered or injected into the spin-down(-up) band. This non-equilibrium spin injection, however, leaves the charge current, equation (\ref{chargecurr}), time independent. The charge current's steady-state solution follows directly from the spin-independent trace over particles and holes in equation (\ref{chargecurr}) and the fact that the transformation matrix $\hat {\cal {U}}(t)$ leaves the diagonal terms of the propagators $\hat{g}^{i/o,<}_{\alpha}$ time independent. The spin current, on the other hand, has components which include the off-diagonal elements of the propagators $\hat{g}^{i/o,<}_{\alpha}$. The off-diagonal elements, which are proportional to the Pauli spin matrices $\sx$ and $\sy$, are time dependent since these are affected by the time dependence of the transformation matrix $\hat {\cal {U}}(t)$. Thus, for a finite tilt angle, $\vartheta\neq 0$, the spin current may be time dependent.

\section{Results}

The precessing spin introduces a new energy scale and the superconducting correlation functions can be expected to be modified due to the scattering processes the precessing spin gives rise to. The starting point is the s-wave superconducting leads and their the spin-singlet pairing amplitudes, $f_s\sim \langle (\psi_{k\uparrow}\psi_{-k\downarrow}-\psi_{-k\downarrow}\psi_{k\uparrow})
\rangle$. When electron- and hole-like quasiparticles interfere  constructively, sharp states inside the superconducting gap called Andreev states are formed. In regular Josephson junctions without magnetically active interfaces, these Andreev states come in degenerate pairs $
\hat\psi_{\uparrow(\downarrow)}$, which can be described by the spinors $\hat \psi_{\uparrow(\downarrow)}=(\psi_{\uparrow(\downarrow)},\psi^\dagger_{\downarrow(\uparrow)})^T$. Each member 
of the spinor, $\psi_{\uparrow(\downarrow)}$, is subjected to transmission $V_{LR(RL)}$ and Andreev retroreflection 
$A_{R(L)}$ and the Andreev bound states are formed when the processes lead to constructive interference along 
closed loops schematically illustrated as $\psi_{L\uparrow(\downarrow)}(k,\varepsilon) 
\stackrel{V_{LR}}{\longrightarrow} \psi_{L\uparrow(\downarrow)}(k,\varepsilon)
\stackrel{A_{R}}{\longrightarrow}  \psi^\dagger_{R\downarrow(\uparrow)}(-k,-\varepsilon)
\stackrel{V_{RL}}{\longrightarrow} \psi^\dagger_{R\downarrow(\uparrow)}(-k,-\varepsilon) 
\stackrel{A_{L}}{\longrightarrow}  \psi_{L\uparrow(\downarrow)}(k,\varepsilon)$. A schematic picture of the scattering processes is
shown in the upper panel of figure \ref{Scattering processes}. The tunneling processes described above are captured by the hopping element
\begin{equation}
\hat v^d = \left(\begin{array}{cc}
 v_0 + v_S \cos{\vartheta} \sz  & 0  \\
0  &   v_0 + v_S \cos{\vartheta} \sz
\end{array}\right) 
\end{equation}
which lets quasiparticles tunnel across the junction with their spin directions and energies unaffected but with different hopping amplitude for spin-up and spin-down. This scattering behavior is captured by the matrix $\hat{v}^{d} \circ \hat{g}_R \circ \hat{v}^{d}$ (see below). The spin-flip part of the hopping element,
\begin{equation}
\hat v^{\uparrow(\downarrow)}=\left(\begin{array}{cc}
v_S \sin{\vartheta} \sigma_{+(-)} & 0 \\
0 &  v_S \sin{\vartheta} \sigma_{-(+)}
\end{array}\right) 
\end{equation}
flips a spin-down(-up) quasiparticle into a spin-up(-down) quasiparticle while changing its energy by $\omega_L$($-\omega_L$).
Here, we have defined $\sigma_{+(-)}=\frac{1}{2}(\sigma_x\pm i\sigma_y)$. This kind of tunneling creates spin-flip processes, e.g.
$\hat\psi_{L\uparrow(\downarrow)}(k,\varepsilon) \stackrel{v^{\downarrow(\uparrow)}_{LR}}{\longrightarrow} \hat\psi_{L\downarrow(\uparrow)}(k_-(k_+),\varepsilon_-(\varepsilon_+))
\stackrel{A_{R}}{\longrightarrow}  \hat\psi^\dagger_{R\uparrow(\downarrow)}(-k_-(-k_+),-\varepsilon_-(-\varepsilon_+))
\stackrel{v^{\uparrow(\downarrow)}_{RL}}{\longrightarrow} \hat\psi^\dagger_{R\downarrow(\uparrow)}(-k,-\varepsilon) 
\stackrel{A_{L}}{\longrightarrow}  \hat\psi_{L\uparrow(\downarrow)}(k,\varepsilon)$, which is a process captured by the elements $\check v^{\uparrow/\downarrow}\!\circ\!\check g^0_R\!\circ\!\check v^{\downarrow/\uparrow}$. 

Focusing on the left side of the interface and parameterizing the matrix $\check\Gamma_{L}$ according to equation (\ref{Xmatrix}), leads one to conclude that the spin-preserving component of $\check\Gamma_{L}$ is a combination of the two processes described above, $\check\Gamma_L^d=\check v^d\!\circ\!\check g^0_R\!\circ\!\check v^d+ \check v^{\uparrow}\!\circ\!\check g^0_R\!\circ\!\check v^{\downarrow} +\check v^{\downarrow}\!\circ\!\check g^0_R\!\circ\!\check v^{\uparrow}$. There are also mixed tunneling processes where tunneling with and without spin flip are combined into the terms $\check\Gamma_L^{\uparrow}=\check v^{\uparrow}\!\circ\!\check g^0_R\!\circ\!\check v^{d}+\check v^{d}\!\circ\!\check g^0_R\!\circ\!\check v^{\uparrow}$ and $\check\Gamma_L^{\downarrow}=\check v^{\downarrow}\!\circ\!\check g^0_R\!\circ\!\check v^{d}+\check v^{d}\!\circ\!\check g^0_R\!\circ\!\check v^{\downarrow}$. These two terms generate a net spin-flip for quasiparticles tunneling across the nanomagnet.

The angle $\vartheta$ between the spin $\SMM$ and the magnetic field $\Heff$ determines the amount of spin-flip 
scattering ranging from zero for parallel alignment to maximum in the case of $\SMM\perp\Heff$. The frequency 
$\omega_L$ sets the amount of energy exchanged between a quasiparticle and the rotating spin during a 
spin-flip event as indicated in figure \ref{Scattering processes} (b). First, we will look at the consequences for the density of states and the charge current due to the scattering caused by the precessing spin. After that, we will take a closer look at the effects on the superconducting pair correlations before we turn to the spin scattering states and the spin current as well as their implications for the leads.

\subsection{Charge currents}

This section reviews some of the results presented in reference [\onlinecite{teber2010}]. Here, however, the charge current results are described in the t-matrix formulation and are included for completeness.

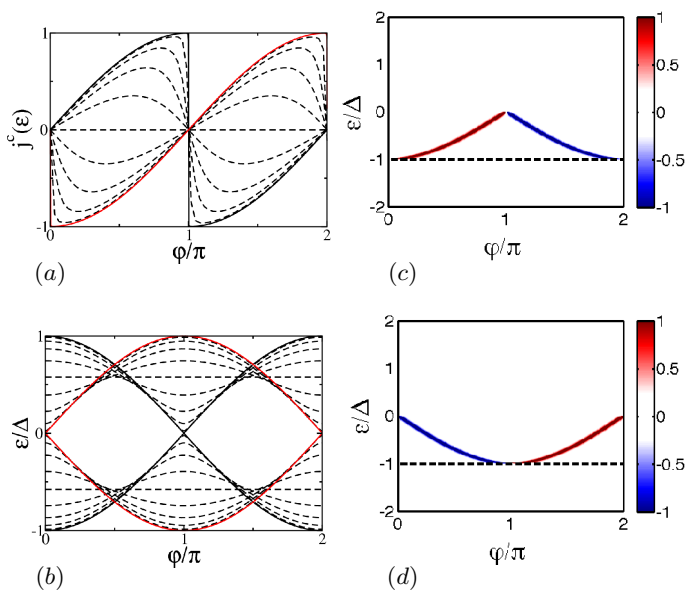
\begin{figure}
  \begin{tabular}{cc cc}
 \scalebox{0.17}{\includegraphics{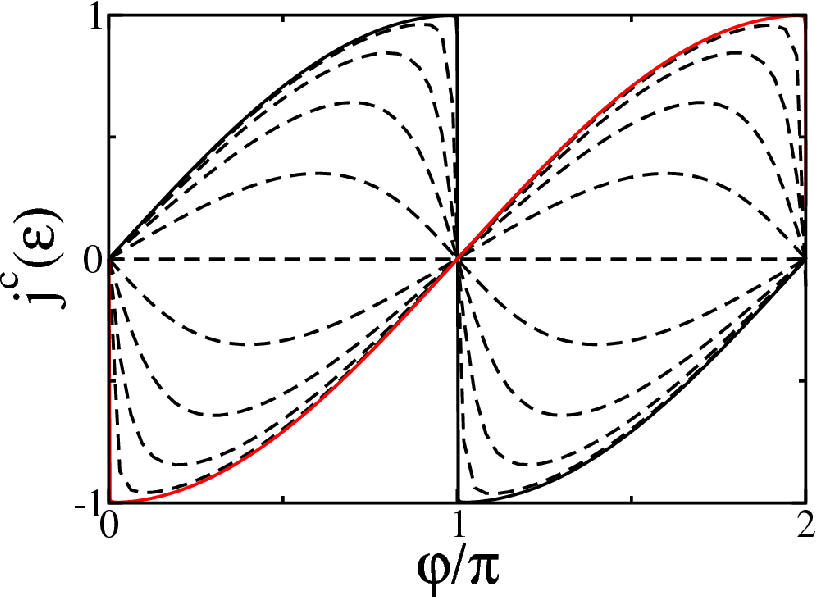}}  & \scalebox{0.32}{\includegraphics{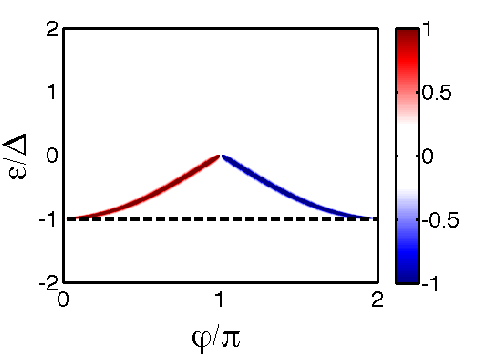}}  \\
  \begin{picture}(100,0)
\put(5,5){\makebox(0,3){$(a)$}}
\end{picture} & \begin{picture}(100,0)
\put(5,5){\makebox(0,3){$(c)$}}
\end{picture} \\
\scalebox{0.17}{\includegraphics{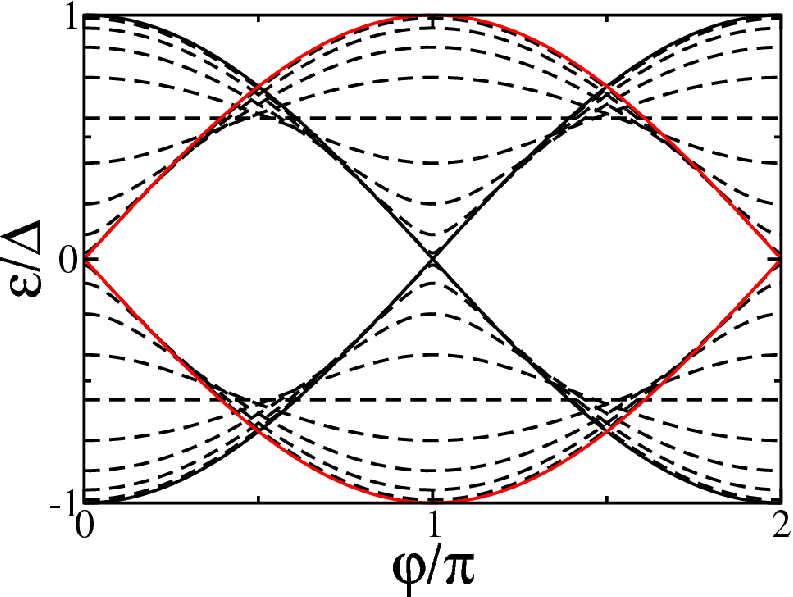}}   & \scalebox{0.32}{\includegraphics{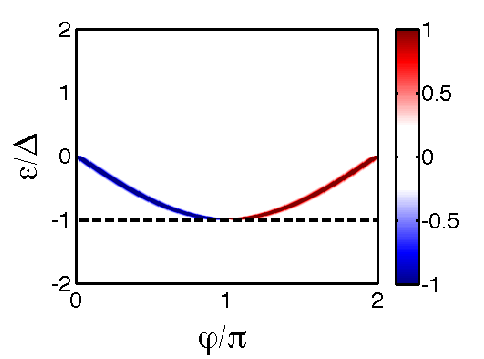}} \\
 \begin{picture}(100,0)
\put(5,5){\makebox(0,3){$(b)$}}
\end{picture} & \begin{picture}(100,0)
\put(5,5){\makebox(0,3){$(d)$}}
\end{picture} \\
   \end{tabular}
\caption{\label{current-phase, 0 to pi}(Color online) Current-phase relation (a) and Andreev levels (b) for a junction whose hopping elements are parameterized as $v_0=\cos{\alpha}$ and $v_S=\sin{\alpha}$ and $\alpha$ is varied from $0$ (solid black line) to $ \pi / 2$ (solid red line) in steps of $0.1 \pi / 2$. The charge current is plotted in units of $e\Delta/\hbar$ for temperature $T\rightarrow 0$.
The charge-current kernel $j^{c,<} (\varepsilon, \varphi)$ shows the population of the Andreev levels as well as the direction in which the populated Andreev levels carry current. 
In (c), $j^{c,<}$ is plotted for
$\alpha=0$. The lower Andreev level, which has energies below the Fermi surface, is populated and carries current in the positive direction (red) for phase differences $0 < \varphi < \pi$ and in the negative direction (blue) for $\pi < \varphi < 2 \pi$. 
In (d), $j^{c,<}$ 
is similarly plotted for $\alpha=\pi/2$. In this case, the populated Andreev level carries a negative current for $0 < \varphi < \pi$ and a positive one for $\pi < \varphi < 2 \pi$. The lower gap edge is indicated by the dashed black line.}
\end{figure}

\subsubsection{Static spin}\label{sec: static spin}

For a static spin with precession frequency $\omega_L=0$, the t-matrix equation (\ref{ttmatrix}) reduces to 
\begin{equation}
[\check{1} - \check{A}^d_{0}]\, \check{t}^d=\check{\Gamma}^d
\end{equation}
which can easily be solved analytically since the spin-up and spin-down bands separate into two sets of equations. Straight-forward calculations of the density of states show that Andreev levels 
form within the superconducting gap and are located at
energies given as
\begin{equation}\label{ABS static spin}
\varepsilon_J = \pm \Delta \sqrt{1-{\cal D}_0 \sin^2{\frac{\varphi}{2}} -{\cal D}_S \cos^2{\frac{\varphi}{2}} }
\end{equation}
where we have defined ${\cal D}_0 = 4 v_0^2/(1+2(v_0^2+v_S^2) +(v_0^2-v_S^2)^2  )$ and a spin-transmission coefficient ${\cal D}_S = 4 v_S^2/(1+2(v_0^2+v_S^2) +(v_0^2-v_S^2)^2  )$. The charge current density is related to the energy dispersion and the occupation
function as \cite{shumeiko97,beenakker91}
\begin{equation}\label{current-ABS rel}
j^c_\alpha=\frac{2e}{\hbar}\frac{\partial \varepsilon_J}{\partial \varphi} \tanh{( \frac{ \varepsilon_J }{2T} )},
\end{equation}
which, given the energy dispersion in equation (\ref{ABS static spin}), is evaluated to
\begin{equation}
j^c_\alpha = \frac{e}{\hbar} \frac{({\cal D}_0-{\cal D}_S) \Delta \sin{\varphi}}{\sqrt{1-{\cal D}_0 \sin^2{( \varphi / 2)}-{\cal D}_S \cos^2{( \varphi / 2)}}}  \tanh{( \frac{ \varepsilon_J }{2T} )}.
\end{equation}
showing that he critical current is reduced due to the spin-flip scattering generated by the embedded nanomagnet \cite{kulik1966}.

If only spin-independent tunneling is present and the spin-dependent hopping strength $v_S\rightarrow0$, the spin-transmission coefficient ${\cal D}_S\rightarrow0$ while ${\cal D}_0\rightarrow {\cal D} = 4 v_0^2/(1+v_0^2)^2$ which is the usual transparency of a Josephson junction. The Andreev levels are now $\varepsilon_J = \pm \Delta \sqrt{1-{\cal D} \sin^2{\frac{\varphi}{2}} }$ and carry a charge current 
\begin{equation}
j^c_\alpha = \frac{e}{\hbar} \frac{{\cal D} \Delta \sin{\varphi}}{\sqrt{1-{\cal D} \sin^2{( \varphi / 2)}}}  \tanh{( \frac{ \varepsilon_J}{2T} )} .
\end{equation}
Both of these relations are shown as solid black lines in figures \ref{current-phase, 0 to pi} (a) and (b). As can be seen in these figures, the junction is in a $0$ state corresponding to the junction's energy being minimized for the phase difference $\varphi=0$.
Increasing the hopping strength of the spin-dependent tunneling until the spin-dependent tunneling dominates, $v_S>v_0$, causes the junction to shift from being in the $0$ state to being a $\pi$ state, as can be seen in figure \ref{current-phase, 0 to pi}(a). When spin-dependent tunneling dominates, the junction's ground state is such that the coupled superconductors have an internal phase shift of 
$\pi$, as predicted in reference [\onlinecite{bulaevskii77}]. Such $\pi$ states can also be observed in junctions where the spin-active barrier has been extended to a ferromagnetic region \cite{buzdin82}. In such junctions, the width of the ferromagnetic layer as well as the strength of the exchange field determine the transport properties. 

If, on the other hand, the spin-independent tunneling is decreased to 0, leaving only spin-dependent tunneling, $(v_0=0, v_S=1)$, the Andreev levels are shifted by $\pi$ to give $\varepsilon_J = \pm \Delta \sqrt{1-{\cal D}_S \cos^2{\frac{\varphi}{2}} }$.
The corresponding charge current is then
\begin{equation}
j^c_\alpha = -\frac{e}{\hbar} \frac{{\cal D}_S \Delta \sin{\varphi}}{\sqrt{1-{\cal D}_S \cos^2{( \varphi / 2)}}}  \tanh{( \frac{ \varepsilon_J }{2T} )}.
\end{equation}
The cross-over from the $0$ to $\pi$ state occurs at $v_0=v_S$ where the Andreev levels are $\varepsilon_J = \pm \Delta/ \sqrt{3}$. 
These Andreev levels are independent of the phase difference between the two superconductors leading to a zero Andreev current consistent with equation (\ref{current-ABS rel}).

In figures \ref{current-phase, 0 to pi} (c) and (d), the current kernel for the left side of the interface $j^{c,<}_{L} (\varepsilon, \varphi)$ is plotted for $v_0=1, v_S=0$ and $v_0=0, v_S=1$, respectively. The current kernel $j^{c,<}_{L} (\varepsilon, \varphi)$ is integrated over energy $\varepsilon$ to give the total current for a specified phase difference, $j^{c}_{\alpha}(t) =  \int \frac{d \varepsilon}{2 \pi}  j^{c,<}_{\alpha} (\varepsilon, \varphi)$.
The kernel $j^{c,<} (\varepsilon, \varphi)$ indicates which states, i.e. Andreev levels and continuum states, are participating in transporting current through the junction. The direction of the current is given by the sign of $j^{c,<}$. As can be seen in panel (c), the lower of the two Andreev levels is occupied which is consistent with quasiparticle states below the Fermi surface being occupied. The same is true for $j^{c,<}$ in panel (d), although the current kernel has been shifted by $\pi$.

\subsubsection{$\pi$ junction with a small tilt angle} \label{section:small theta charge current}

For a junction with zero tilt angle, the solution of the boundary condition problem reduces to the static spin case. The t-matrix equation is simply 
\begin{equation}
\check t^0_\alpha(t,t^\prime)=\check \Gamma^0_\alpha(t,t^\prime) +\lbrack \check \Gamma^0_\alpha\!\circ\!\check g^{0}_\alpha \!\circ\!  \check t^0_\alpha\rbrack (t,t^\prime)
\end{equation} 
with $\check \Gamma^0_\alpha = \check \Gamma^d_\alpha = \lbrack \check v^d \circ \check g^{0}_\alpha \circ \check v^d \rbrack (t,t^\prime)$. Since the hopping elements are time independent, the t-matrix equation can be transformed into energy space where the solution can be found as
\begin{equation}
\check t^0_\alpha (\varepsilon) =\lbrack \lbrack \check 1-\check \Gamma^0_\alpha\!\circ\!\check g^{0}_\alpha  \rbrack^{-1}\!\circ\!\check \Gamma^0_\alpha \rbrack (\varepsilon). 
\end{equation} 
The t matrices are then used to calculate the incoming and outgoing propagators according to equation (\ref{boundary}). The poles of these propagators subsequently give the Andreev levels, $\varepsilon^0_J$, which turn out to be the same as for a static spin ($\omega_L=0$) with arbitrary tilt angle.

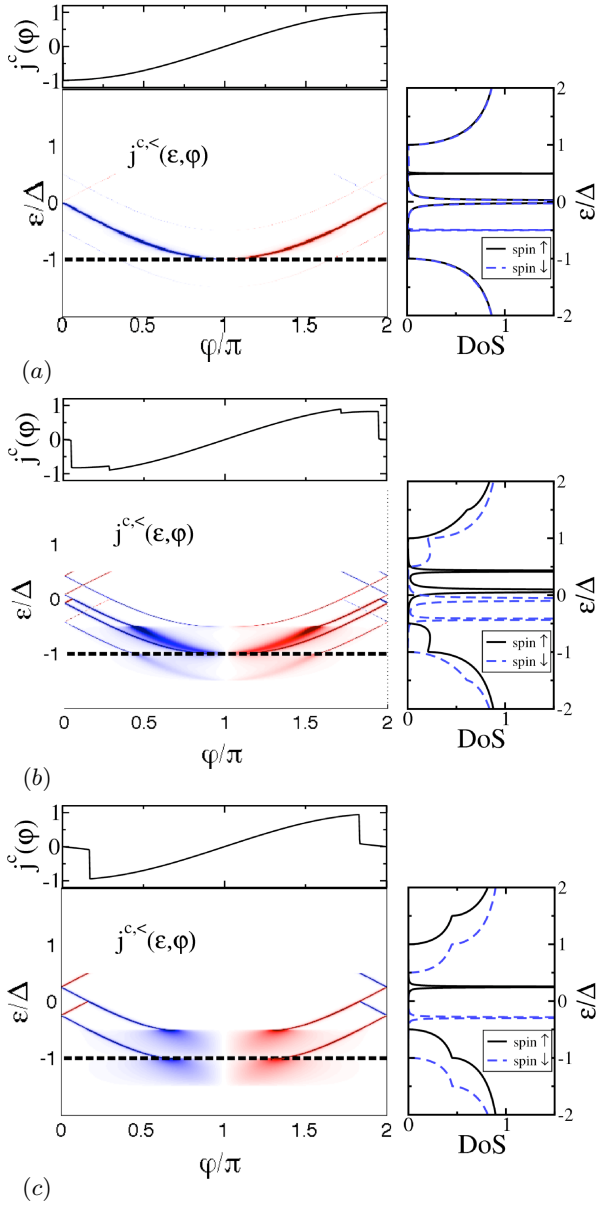
\begin{figure}
  \begin{tabular}{cc cc}
\scalebox{0.11}{\includegraphics{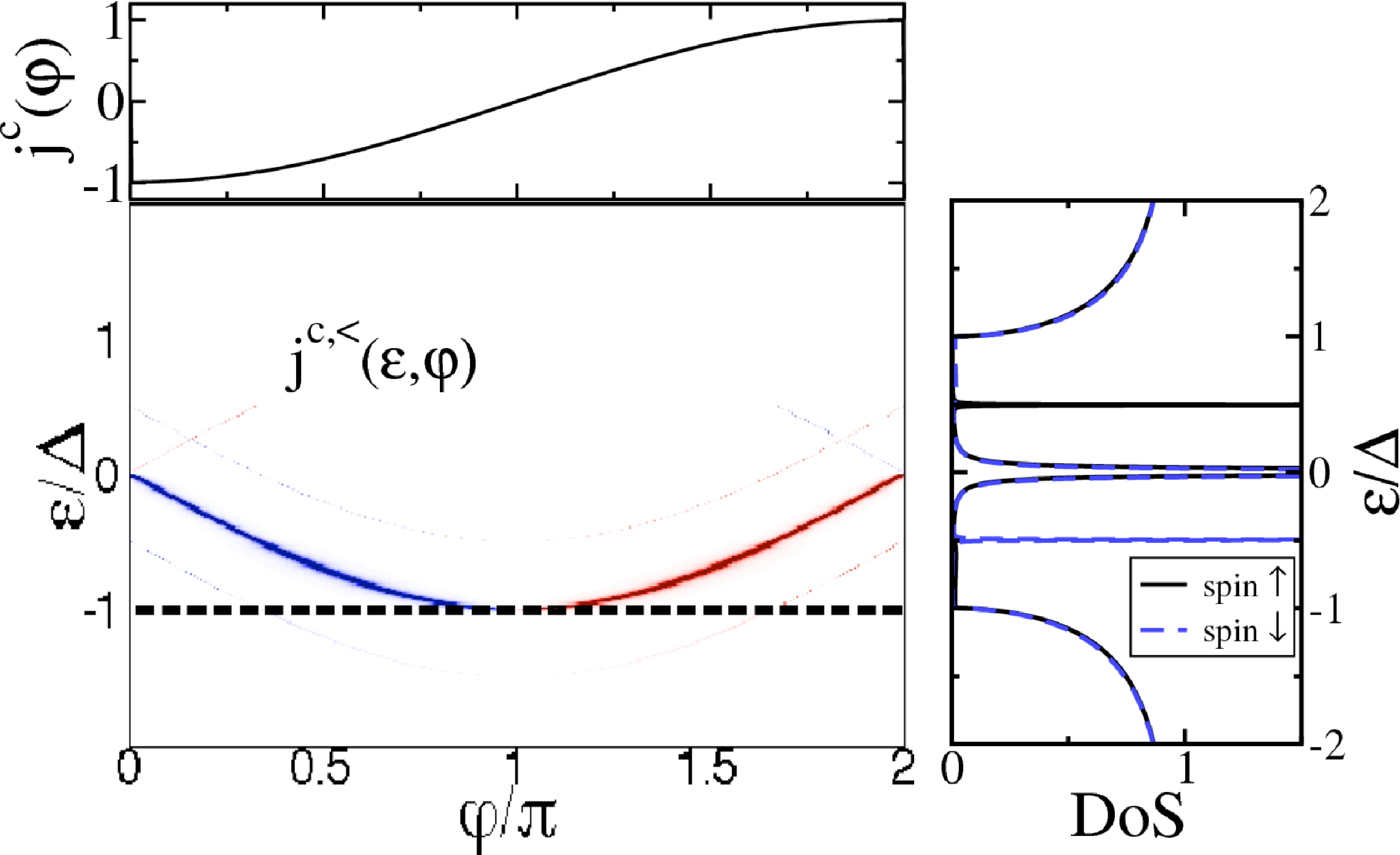}}   \\
\begin{picture}(220,0)
\put(10,5){\makebox(0,3){$(a)$}}
\end{picture}\\
\scalebox{0.11}{\includegraphics{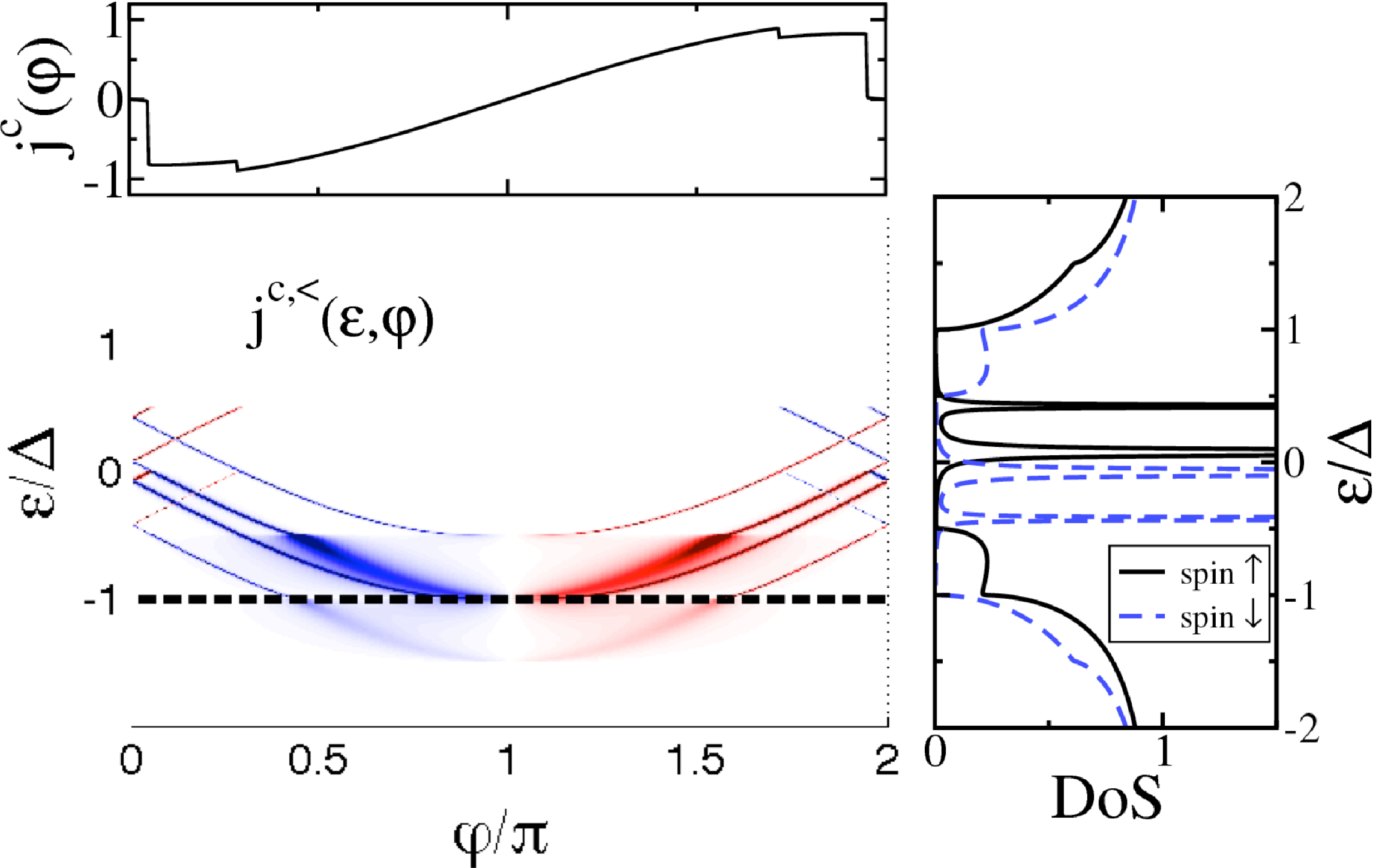}} \\
\begin{picture}(220,0)
\put(10,5){\makebox(0,3){$(b)$}}
\end{picture}\\
\scalebox{0.11}{\includegraphics{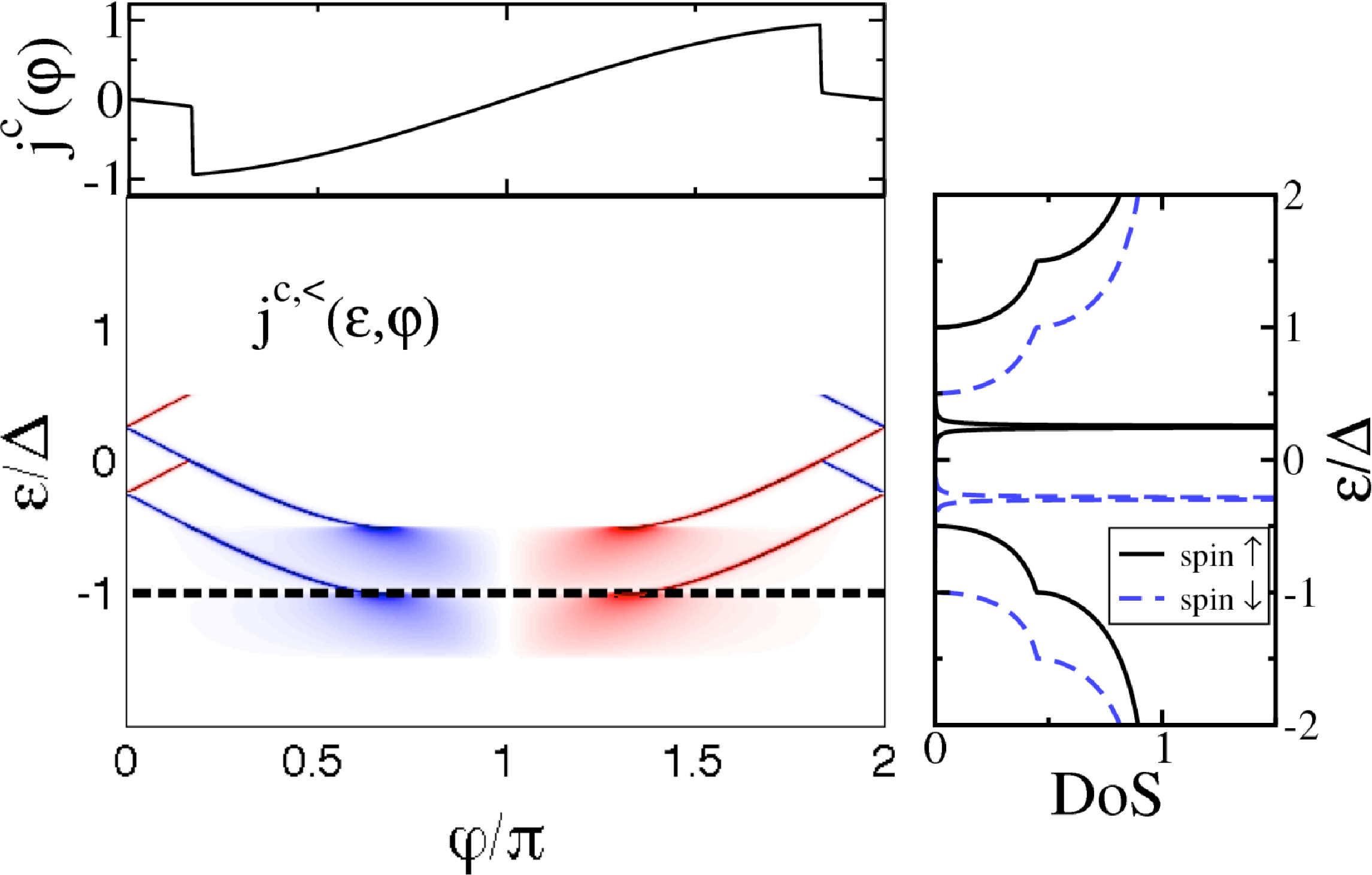}}\\
\begin{picture}(220,0)
\put(10,5){\makebox(0,3){$(c)$}}
\end{picture}
  \end{tabular}
\caption{\label{spectralcurrentplots}
(Color online) The charge current-phase relation, $j^c(\varphi)$, charge current kernel, $j^{c,<} (\varepsilon, \varphi)$, and density of states are plotted for a $\pi$ junction ($v_0=0, v_S=1$) with tilt angle (a) $\vartheta=0.1\pi/2$, (b) $\vartheta=\pi/4$ and (c) $\vartheta=\pi/2$. The temperature of the leads is put to $10^{-5}\Delta$. The spectral charge current, $j^{c,<} (\varepsilon, \varphi)$, shows the Andreev levels and their population.  
At some phase differences $\varphi_c<\varphi<2\pi-\varphi_c$, scattering between the Andreev levels and continuum states, which are indicated with the black dashed line, cause a broadening of the Andreev levels. The charge current, $j^{c} (\varphi)$ (plotted in units of $e\Delta/\hbar$), is the energy-integrated spectral current and displays abrupt jumps at phase differences where Andreev levels become populated/unpopulated. The density of states (DoS) on the right sides are plotted for the phase difference $\varphi=0$ and shows the splitting of the spin-up and spin-down Andreev levels as well as the scattering of the continuum levels into the gap. The precessing spin has a frequency of $\omega_L=0.5\Delta$ in all plots.}
\end{figure}

If the classical spin acquires a small tilt angle $\vartheta = \delta$ such that $\sin\delta\approx\delta$ and precesses around the $\ez$ axis with frequency $\omega_L$, the spin-dependent hopping element may be approximated with $v_S \eS(t)\cdot \sv =v_S \big ( \sz+ \delta\, {\rm e}^{-i \omega_L t \sz}\sx\big)$ assuming that $v_0=0$. This means that a quasiparticle may either tunnel across the junction with its spin and energy conserved, or it may gain or lose energy $\omega_L$ while having its spin reversed as illustrated in figure \ref{Scattering processes} (b).
Then, to first order in $\delta$, the t matrices are $\check t_\alpha = \check t^0_\alpha + \delta \check t_\alpha$ while $\check \Gamma_\alpha = \check \Gamma^0_\alpha + \delta \check \Gamma_\alpha$. Note that $\delta \check \Gamma_\alpha = e^{-i \omega_L t} \delta \check \Gamma_\alpha^\uparrow + e^{i \omega_L t} \delta \check \Gamma_\alpha^\downarrow$ and has no diagonal time-independent term. Inserting this into the t-matrix equation and disregarding terms of order $\delta^2$ and higher gives an equation for $\delta \check t_\alpha$ as
\begin{eqnarray}
\delta \check t_\alpha(t,t^\prime)&=& \lbrack \delta \check \Gamma_\alpha \circ (\check 1 + \check g^{0}_\alpha \circ \check t^0_\alpha) \rbrack (t,t^\prime) \\ \nonumber
& &+\lbrack \check \Gamma^0_\alpha\!\circ\!\check g^{0}_\alpha \!\circ\! \delta \check t_\alpha\rbrack (t,t^\prime).
\end{eqnarray}
Since there are no energy shifts in the last term, the t-matrix equation can be written according to equation (\ref{ttmatrix}) as
\begin{equation}\left(\begin{array}{ccc}
\check{1} - \check{A}^d_{1}& 0 & \check{0} \\
0 & \check{1} - \check{A}^d_{0} & 0  \\
\check{0} & 0 &\check{1} - \check{A}^d_{-1} 
\end{array}\right)
\left(\begin{array}{c}
\delta \check{t}^\uparrow_\alpha \\ \delta \check{t}^d_\alpha\\ \delta \check{t}^\downarrow_\alpha \end{array}\right)
=\left(\begin{array}{c} 
\delta \check{\Gamma}^{\prime\uparrow}_\alpha  \\ 0 \\ \delta \check{\Gamma}^{\prime\downarrow}_\alpha  
\end{array}\right)
\label{deltattmatrix}
\end{equation}
where $\delta \check{\Gamma}^{\prime\uparrow/\downarrow}_\alpha=\delta \check{\Gamma}^{\uparrow/\downarrow}_\alpha\!\circ\!\, (1+\check g_\alpha\!\,\circ\!\,\check t^0_\alpha ) $.
The time-independent change in the t matrix is consequently zero and the total t matrix is given by $\check t_\alpha=\check t^0_\alpha +e^{-i \omega_L t} \delta \check t_\alpha^\uparrow+e^{i \omega_L t}\delta \check t_\alpha^\downarrow$. In other words, a small tilt angle adds two time-dependent components to the t matrix. These two components have poles at $\varepsilon^0_J \pm \omega_L$ since quasiparticles gain or lose energy $\omega_L$ while being spin flipped. However, these sidebands do not show up in the Andreev levels for the charge current in this first-order approximation since $t_\alpha^d = t_\alpha^0$. When the tilt angle becomes large enough for higher-order processes to become important, the sidebands do appear in the Andreev level spectrum which is shown explicitly in
figure \ref{spectralcurrentplots}. Spin-down (spin-up) quasiparticles in the state $\varepsilon^0_J$ are scattered into the upper (lower) sideband, which has an opposite spin direction and the quasiparticles exchange the energy $\omega_L$ in this spin-flip process with 
the precessing spin. A cross-section of the spin-resolved density of states for $\varphi=0$ is shown in figure \ref{spectralcurrentplots}(a).

In addition to the sidebands, the scattering processes also lead to a nonequilibrium population of states. The quasiparticles occupying the lower Andreev level $\varepsilon^0_J$ are scattered into states with energies larger than the Fermi energy, leading to an occupation of states above the Fermi surface. Quasiparticles are also scattered into the continuum states below the gap edge. The continuum scattering leads to a reduced life-time for the quasiparticles in the state $\varepsilon^0_J$ and can be seen as a broadening of the state for energies $\varepsilon^0_J\le-\Delta+\omega_L\Delta$. Despite the time-dependent dynamics of the rotating spin and the nonequilibrium population of states, the charge current is still time independent consistent with reference [\onlinecite{zhu2003}].

\subsubsection{$\pi$ junction with arbitrary tilt angle}

As the tilt angle increases, scattering into the sidebands increases. The spin-degenerate state $\varepsilon^0_J$ now splits up into a 
spin-up and a spin-down state and the spin-flip scattering still occurs between states separated by energy $\omega_L$.  The population of the sharp states, the Andreev levels, can be understood on the basis of the spin-flip scattering processes. As before, the spin-flip scattering connects two states in opposite spin-bands separated by energy $\omega_L$. This connection results in a population of the upper state under the condition that the lower state is occupied. In other words, if the spin-down propagator is occupied, the spin-up propagator is occupied as well. Vice versa, if the spin-down propagator is not occupied, the spin-up propagator is not occupied either.
This non-equilibrium occupation is an effect of the precessing classical spin which scatters between the two spin states. 
The fine details of this are shown in figure \ref{spectralcurrentplots} where we plot $j^{c,<}(\varepsilon;\varphi)$ for some values of the tilt angle $\vartheta$.

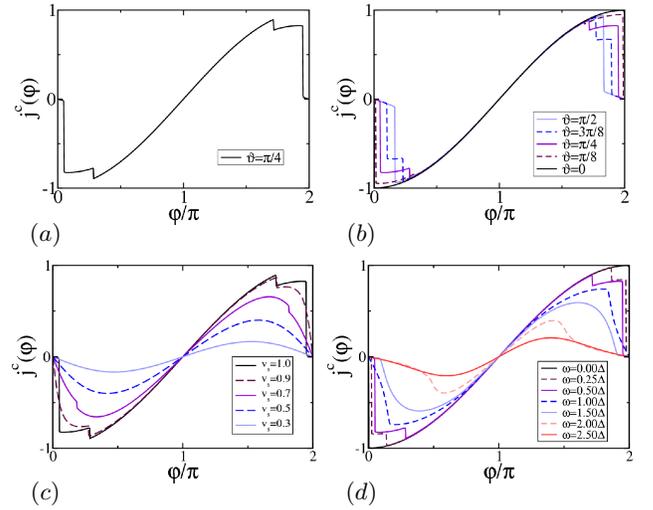
\begin{figure}
 \begin{tabular}{cc cc}
 \scalebox{0.16}{\includegraphics{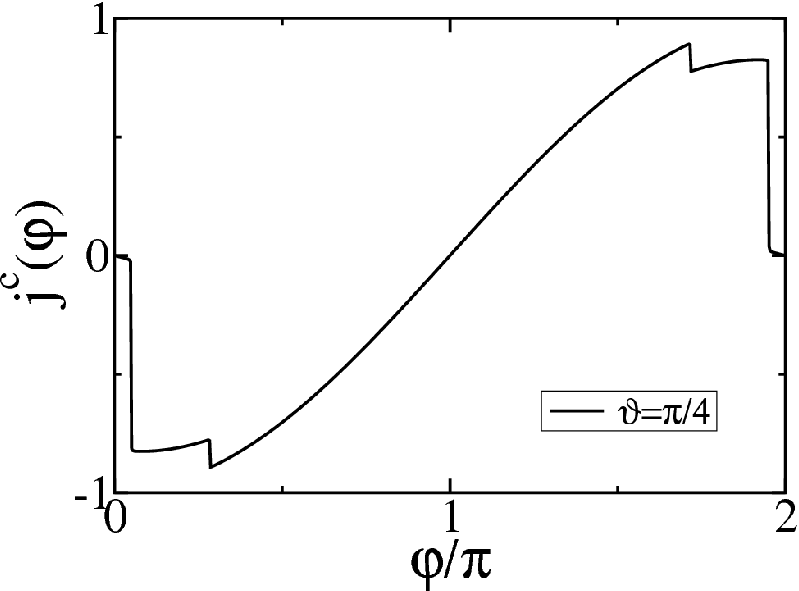}}  &   \scalebox{0.16}{\includegraphics{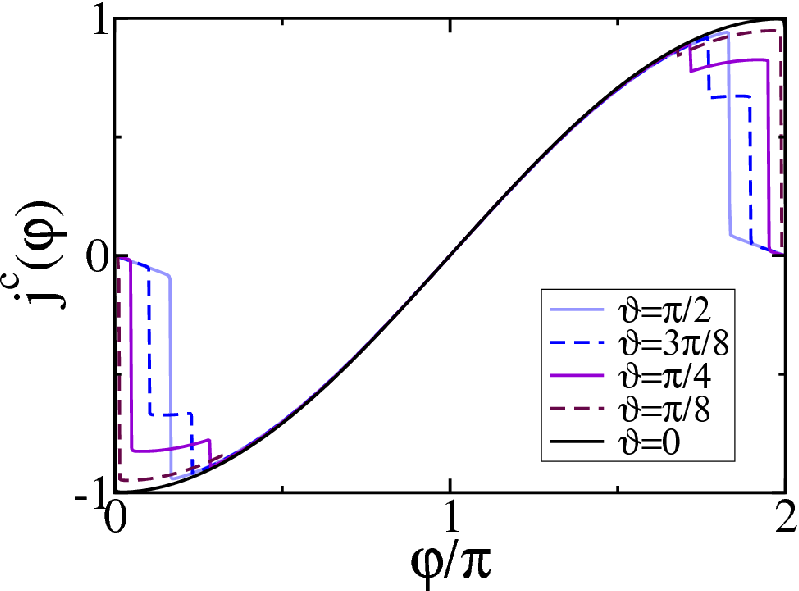}} \\
   \begin{picture}(100,0)
\put(5,5){\makebox(0,3){$(a)$}}
\end{picture} & \begin{picture}(100,0)
\put(5,5){\makebox(0,3){$(b)$}}
\end{picture} \\
      \scalebox{0.16}{\includegraphics{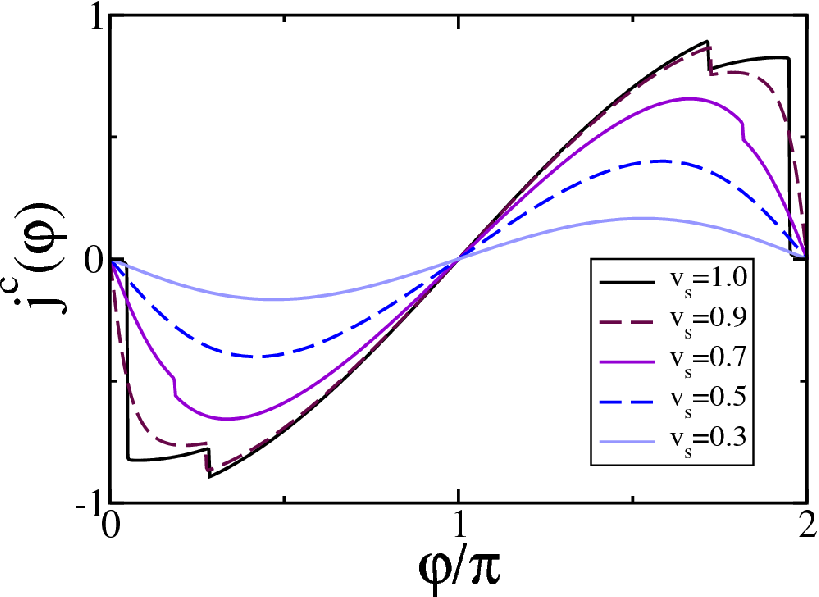}}   &     \scalebox{0.16}{\includegraphics{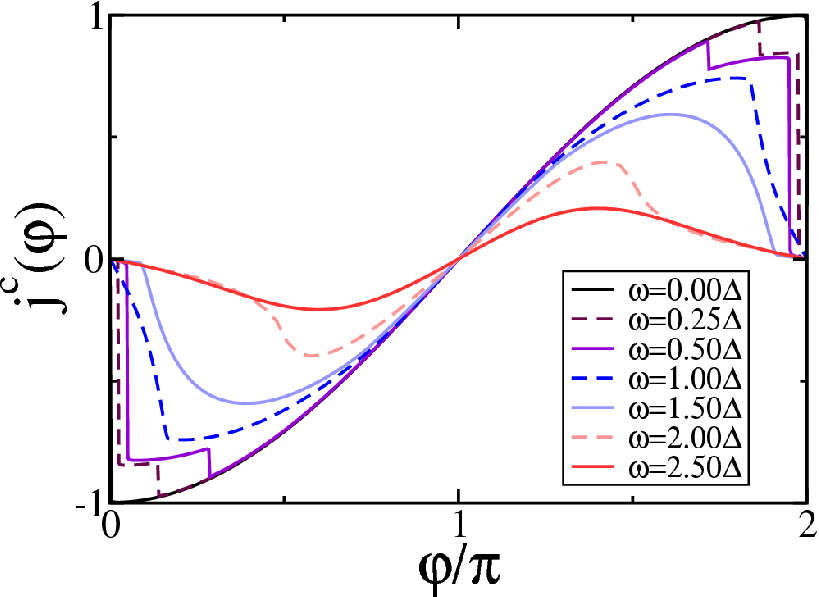}} \\
        \begin{picture}(100,0)
\put(5,5){\makebox(0,3){$(c)$}}
\end{picture} & \begin{picture}(100,0)
\put(5,5){\makebox(0,3){$(d)$}}
\end{picture} \\
  \end{tabular}
\caption{\label{charge current-phase}
(Color online) (a) Current-phase relation for junction with $(v_0=0,v_S=1)$, $\omega = 0.5 \Delta$ and $\vartheta=\pi/4$.  (b) 
Current-phase relations for $(v_0=0,v_S=1)$, $\omega_L=0.5 \Delta$, and $\vartheta$ varies between $0$ and $\pi/4$. (c)
$\omega=0.5\Delta$, $\vartheta=\pi/4$, $v_0=0$, and $v_S$ varies between $1.0$ and $0.3$ (d) $(v_0=0,v_S=1)$, $\vartheta=\pi/4$ and $\omega_L$ varies between $0.00\Delta$ and $2.50\Delta$. As seen sharp step-like features in the current-phase relation may be
washed out by either having a limited transparency $v_S\ll1$ or by having a large frequency $\omega_L\gtrsim\Delta$. In all four panels the
temperature is set to zero and the currents are plotted in units of $e\Delta/\hbar$.}
\end{figure}

This nonequilibrium occupation of states generates a current-phase relation quite different from the non-spin-flip current-phase relation
as shown in figure \ref{charge current-phase}.
In the laboratory frame, the tilt angle $\vartheta$ between the precessing spin and the external magnetic field determines the splitting of the Andreev levels. The splitting in turn determines the current-phase relation and the locations of the abrupt jumps as can be seen in panel \ref{charge current-phase} (b). If the classical spin is aligned with the magnetic field, there are no spin-flip scattering processes and the only tunneling processes taking place are the usual ones where the quasiparticles' energies are conserved (as in figure \ref{Scattering processes} (a)). Hence there is no splitting of Andreev levels and there are no abrupt jumps in the current-phase relation. If, on the other hand, the classical spin precesses in the plane, there are only spin-flip scattering processes present. Hence, the Andreev levels split up into a single spin-up and a spin-down levels. Since each spin band only has one Andreev level, there is only one jump in the current-phase relation.

When the spin-dependent hopping strength, $v_s$, is decreased, the abrupt jumps in the current-phase relation disappear as can be seen in panel \ref{charge current-phase} (c). The reason is that the Andreev levels are closer to the gap edges for lower transparencies. 
If they are close enough they even merge with the continuum states that have been scattered into the gap. Similar effects are created when the precession frequency $\omega_L$ is increased. The continuum states are scattered further into the gap, removing the sharp Andreev levels.
This modification of the Andreev levels causes the jumps in the current-phase relation to be smoothed out (panel \ref{charge current-phase} (d)).

\subsubsection{$\pi$ junction with tilt angle $\vartheta=\pi/2$}

\begin{figure}
  \begin{tabular}{cc}
  \scalebox{0.115}{\includegraphics{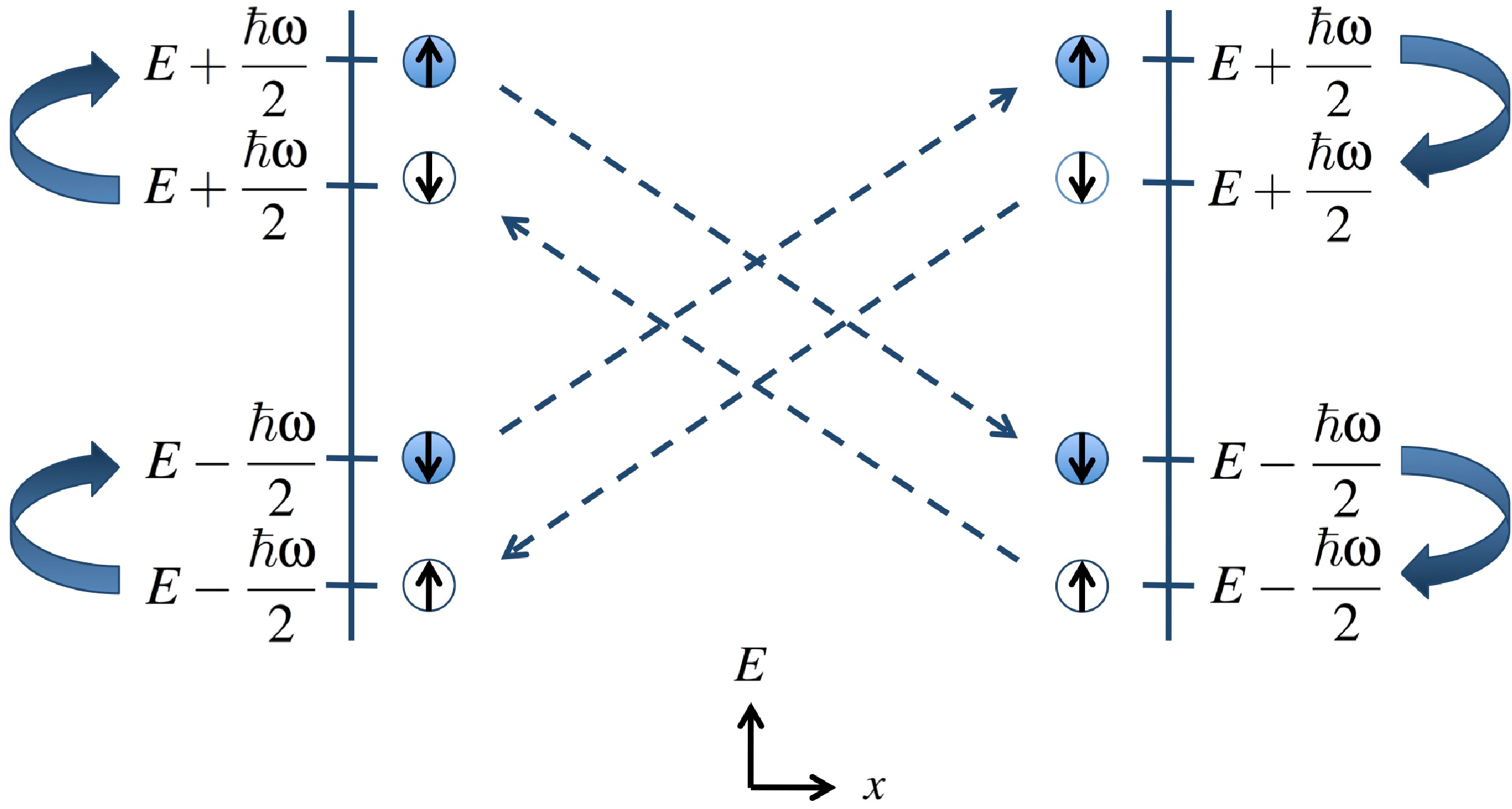}}  
  \end{tabular}
\caption{\label{Scattering processes, plane}
(Color online) Scattering processes in a Josephson junction coupled to a spin precessing with tilt angle $\vartheta=\pi/2$ and $v_0=0$. The tunneling processes are such that all quasiparticles traversing the junction reserve their spins while gaining or losing energy $\omega_L$.}
\end{figure}

In the limit $\vartheta \rightarrow \pi/2$ in which the classical spin precesses in the plane, the only spin-dependent tunneling present generates spin-flip scattering processes causing quasiparticles gain or lose energy $\omega_L$ as shown in figure \ref{Scattering processes, plane} (see also references \cite{holmqvist09,teber09}). Further assuming that the spin-independent tunneling is negligible, i.e. $v_0=0$, the hopping element is 
\begin{equation}
\hat v =e^{-i \omega_L t} \hat v^{\uparrow}+e^{i \omega_L t} \hat v^{\downarrow}
\end{equation}
where 
\begin{eqnarray}
\hat v^{\uparrow}&=&v_S \sin{\vartheta}  \left(\begin{array}{cc}
\sigma_+ & 0 \\
0 &\sigma_-
\end{array}\right) \\ \nonumber
&\mbox{and}& \\ \nonumber
\hat v^{\downarrow}&=&v_S \sin{\vartheta} \left(\begin{array}{cc}
\sigma_- & 0 \cr
0 & \sigma_+ 
\end{array}\right)
\end{eqnarray}
which leads to a $\check\Gamma_\alpha$ where only the diagonal term is nonzero and has the form 
\begin{equation}
\check\Gamma_\alpha^d=\lbrack \check v^\uparrow \!\circ\! \check g^{0}_{R/L}\!\circ\!\check v^\downarrow \rbrack (t,t^\prime) +\lbrack \check v^\downarrow \!\circ\! \check g^{0}_{R/L}\!\circ\!\check v^\uparrow \rbrack (t,t^\prime).
\end{equation}
Carrying out the convolutions, we find that $\lbrack \check v^\uparrow \!\circ\! \check g^{0}_{R}\!\circ\!\check v^\downarrow \rbrack (t,t^\prime)$ is associated with the spin-up quasiparticles while $\lbrack \check v^\downarrow \!\circ\! \check g^{0}_{R}\!\circ\!\check v^\uparrow \rbrack (t,t^\prime)$ is related to the spin-down quasiparticles. These $\check\Gamma_\alpha$ matrices result in a t-matrix equation given as
\begin{equation}\left(\begin{array}{ccc}
\check{1} - \check{A}^d_{1}& 0 & \check{0} \\
0 & \check{1} - \check{A}^d_{0} & 0  \\
\check{0} & 0 &\check{1} - \check{A}^d_{-1} 
\end{array}\right)
\left(\begin{array}{c}
 \check{t}^\uparrow_\alpha \\ \check{t}^d_\alpha\\  \check{t}^\downarrow_\alpha \end{array}\right)
=\left(\begin{array}{c} 
0 \\ \check\Gamma^d_\alpha \\ 0 
\end{array}\right)
\label{planettmatrix}
\end{equation}
which immediately results in the only nonzero t-matrix term being $\check t^d_\alpha$. These two properties ensure that we can divide the t-matrix equation into two sets of equations, one for each spin band, and Fourier transform the equations into energy space. For the left side, we find
\begin{equation}
\check{t}_{L\spinup}(\varepsilon)=
\bigg\lbrack \frac{1}{|v_s|^2}\check{g}_{R\spindown,\infty}^{-1}(\varepsilon-\omega_L)-
\check{g}_{L\spinup,\infty}(\varepsilon)\bigg\rbrack^{-1}.
\label{efftmatrix}
\end{equation}
The corresponding t matrix for the right side is found by interchanging $L$ and $R$. The spin-down component is given by  substituting $\uparrow\leftrightarrow\downarrow$ and reversing the sign of the energy shift, $-\omega_L\rightarrow +\omega_L$. The form of the t matrix indicates that a Green's function at a given energy $\varepsilon$ on one side of the interface is connected to Green's functions at energies $\varepsilon\pm\omega_L$ on the other side of the interface.

As before, the current is carried by sharp Andreev levels inside the gap as well as continuum states in regions $\pm\omega_L$ around the gap edges $\pm\Delta$ as shown in figure \ref{spectralcurrentplots} (c). The Andreev levels are given by
\begin{eqnarray}
\varepsilon_{J,\spinup} &=& \frac{\omega}{2} \pm \Delta \sqrt{1+ f(\omega,\varphi,v_s)} \\ \nonumber
\varepsilon_{J,\spindown} &=&  - \frac{\omega}{2} \pm \Delta \sqrt{1+f(\omega,\varphi,v_s )  } 
\end{eqnarray}
where 
\begin{eqnarray}
& &f(\omega,\varphi,v_s) =  \frac{8 v_s^4}{(1-v_s^4)^2} \cos^2{\frac{\varphi}{2}} + \left( \frac{\omega}{2 \Delta} \right)^2  \\ \nonumber
& & - \frac{(1+v_s^4)^2}{(1-v_s^4)^2} \sqrt{\frac{4v_s^4}{(1+v_s^4)^2} (1+\cos{\varphi})^2 + 4 \left( \frac{1-v_s^4}{1+v_s^4} \right)^2 \left( \frac{\omega}{2 \Delta} \right)^2 }.
\end{eqnarray} 
In the laboratory frame, the number of Andreev levels present when $\vartheta=\pi/2$ is thus half the number of states in junctions with $0<\vartheta<\pi/2$. The states lie symmetrically around $\pm\omega_L/2$ and disperse with the phase difference $\varphi$ until they touch the gap edges and merge with the continuum at $\varphi_c(\omega_L)$. The sharp states reappear again at $2\pi-\varphi_c(\omega_L)$. The population of the states is indicated in figure \ref{spectralcurrentplots} (c) showing the current kernel $j^{c,<}(\varepsilon,\varphi)$. Since the number of states is reduced, two for the incoming propagator and two for the outgoing propagator, there is only one abrupt jump, instead of two, in the current-phase relation for $\vartheta=\pi/2$ shown in figure \ref{charge current-phase} (b).

\subsubsection{Tunnel limit}

For a low transparency junction with $v_0,v_S\ll 1$, we may use the first-order approximation $\check t_\alpha(t,t^\prime)=\check \Gamma_\alpha(t,t^\prime)$ and obtain an analytic expression for the
charge current. The charge current between the two spin-singlet superconductors is time-independent and is given by the diagonal hopping element $\check t_L^d=\check\Gamma_L^d$. In the absence of an applied voltage bias, the current is given by the anomalous Green's functions and has the form (at zero temperature)
\begin{equation}
j^{c} =
\left\{
        \begin{array}{ll}
                \frac{e \Delta}{2\hbar} \left( ({\cal D}_0 - {\cal D}_S\cos^2\vartheta)  - \frac{2}{\pi}{\cal D}_S\sin^2\vartheta\,K\big(\frac{\om_L}{2 \Delta }\big)\right) \sin \varphi,  \\ 
                \,\,\,\,\,\,\,{\rm for }\,\om_L < 2 \Delta, \\
                \frac{e \Delta}{2\hbar} \left( ({\cal D}_0 - {\cal D}_S\cos^2\vartheta)  - \frac{4\Delta}{\pi \om_L}{\cal D}_S\sin^2\vartheta\,K\big(\frac{\om_L}{2 \Delta }\big)\right) \sin \varphi,  
                \\ \,\,\,\,\,\,\,{\rm for }\,\om_L > 2 \Delta, \\
          \end{array}
\right.
\label{ChargeCurrentTunnel}
\end{equation}
where $K$ is the complete elliptic integral of the first kind. For an alternative derivation, see reference [\onlinecite{teber2010}]. The first term is due to the processes which do not flip the spin of the tunneling electrons, $\check v^{d}\!\circ\!\check g^0_R\!\circ\!\check v^{d}$. The part of the current depending on the precession frequency is entirely due to the spin-flip processes described by $\check v^{\downarrow}\!\circ\!\check g^0_R\!\circ\!\check v^{\uparrow}$ and $\check v^{\downarrow}\!\circ\!\check g^0_R\!\circ\!\check v^{\uparrow}$. Equation (\ref{ChargeCurrentTunnel}) shows that also in the limit of low transparency but arbitrary precession frequency, the critical current is reduced by spin-flip scattering \cite{kulik1966} and even reversed in junctions dominated by spin-flip scattering \cite{bulaevskii77}.

\subsection{Spin currents}
\label{sec: spin currents}

The charge and spin currents in equations (\ref{chargecurr}) and (\ref{spincurr}) are evaluated as the difference between the incoming and outgoing propagators in each lead $\alpha$. Defining a current matrix as $\hat{j}^<_\alpha(\varepsilon,t)=\hat{g}^{i,<}_{\alpha}(\varepsilon,t) - \hat{g}^{o,<}_{\alpha}(\varepsilon,t)$ and using the expressions for the incoming and outgoing propagators, equation (\ref{boundary}), it is easily found that the current matrix is given by the commutator
\begin{equation}\label{spin commutator 1}
\hat{j}^<_\alpha(\varepsilon,t)=2\pi i[\check{t}_\alpha(\varepsilon,t),\check{g}^0_\alpha(\varepsilon)]^<_\circ.
\end{equation}
The t-matrices have the spin properties of equation (\ref{Xmatrix}) and can, without any loss of generality, be divided into a scalar singlet part $\check{t}^s_\alpha=\check{t}^d_\alpha-\check{t}^z_\alpha$ and a spin-vector triplet part $\check t^t_\alpha(\varepsilon,\omega_L)=\check t^z_\alpha(\varepsilon,\omega_L)+{\rm e}^{-i \omega_L t}\check t^\uparrow_\alpha(\varepsilon,\omega_L)+{\rm e}^{i \omega_L t}\check t^\downarrow_\alpha(\varepsilon,\omega_L)$ such that 
\begin{eqnarray}\label{tmatrix, scalar and triplet}
\check t_\alpha(t,t')&=&\int \frac{d \varepsilon}{2 \pi}{\rm e}^{-i\varepsilon(t-t')}\bigg\lbrack 
\check t^s_\alpha(\varepsilon,\omega_L)+\check t^t_\alpha(\varepsilon,\omega_L)\bigg\rbrack . \nonumber
\end{eqnarray}
The scalar and spin-vector parts of the t matrix in Nambu-spin space are, e.g. for the retarded (R) matrix component,
\begin{equation} \label{t singlet}
\hat{t}^{s,R}_\alpha=\left(\begin{array}{cc}\gamma^{s,R}_\alpha & \phi^{s,R}_\alpha i\sy \\ i\sy {\tilde \phi}^{s,R}_\alpha & {\tilde \gamma}^{s,R}_\alpha \end{array}\right)
\end{equation}
and
\begin{equation}\label{t triplet}
\hat{t}^{t,R}_\alpha=\left(\begin{array}{cc} \gamma^{t,R}_\alpha & (\phi^{t,R}_\alpha) i\sy \\ i\sy ({\tilde 
\phi}^{t,R}_\alpha) & - \sigma_y ({\tilde \gamma}^{t,R}_\alpha )\sigma_y  \end{array}\right)
\end{equation}
where $\gamma^{t,R}_\alpha/\phi^{t,R}_\alpha$ and $\tilde{\gamma}^{t,R}_\alpha/\tilde{\phi}^{t,R}_\alpha$ are $2\times2$ matrices in spin space. The spin-triplet t matrix can be parameterized similarly to the Green's function in equation (\ref{scalar and triplet green's function}),
\begin{equation}\label{t triplet xyz}
\hat{t}^{t,R}_\alpha=\left(\begin{array}{cc} \gammavec^{t,R}_\alpha\cdot\svec & (\phivec^{t,R}_\alpha\cdot\svec) i\sy \\ i\sy ({\tilde 
\phivec}^{t,R}_\alpha\cdot\svec) & - \sigma_y ({\tilde \gammavec}^{t,R}_\alpha\cdot\svec )\sigma_y  \end{array}\right),
\end{equation}
although the parameterization using the basis $(z,\uparrow,\downarrow)$, e.g. $\gamma^{t,R}_\alpha=\gamma^{t,R}_{\alpha,z}\sz+e^{-i\omega t}\gamma^{t,R}_{\alpha,\uparrow} \sigma_++e^{i\omega t}\gamma^{t,R}_{\alpha,\downarrow} \sigma_-$, has a more straight-forward time dependence.

From the current matrix, the division of the t matrix into scalar and triplet components, and the fact that $\check{g}_\alpha^0$ is a scalar, it is clear that the spin current is due to the triplet components of the t matrix
\begin{equation}
\jvec^{s}_\alpha(t) =  \frac{1}{4}  \int \frac{d \varepsilon}{8 \pi i} 2\pi i \mbox{Tr} [ \hat{\tau}_3 \hat{\svec} 
 [\check{t}^t_\alpha(\varepsilon,t),\check{g}^0_\alpha(\varepsilon)]^<_\circ ]
\end{equation}
since
\begin{equation}
\int \frac{d}{2\pi} \varepsilon\mbox{Tr} [ \hat{\tau}_3 \hat{\svec} 
 [\check{t}^s_\alpha(\varepsilon,t),\check{g}^0_\alpha(\varepsilon)]^<_\circ ]=0.
\end{equation}
As a passing note, the charge current is due to the scalar component of the t matrix,
\begin{equation}
j^{c}_\alpha(t) =  \frac{e}{2\hbar}  \int \frac{d \varepsilon}{8 \pi i} 2\pi i \mbox{Tr} [ \hat{\tau}_3 
 [\check{t}^s_\alpha(\varepsilon,t),\check{g}^0_\alpha(\varepsilon)]^<_\circ ],
\end{equation}
since
\begin{equation}
\int \frac{d}{2\pi} \varepsilon\mbox{Tr} [ \hat{\tau}_3  
 [\check{t}^t_\alpha(\varepsilon,t),\check{g}^0_\alpha(\varepsilon)]^<_\circ ]=0.
\end{equation}

\subsubsection{Small tilt angle, $\vartheta\ll\pi/2$}

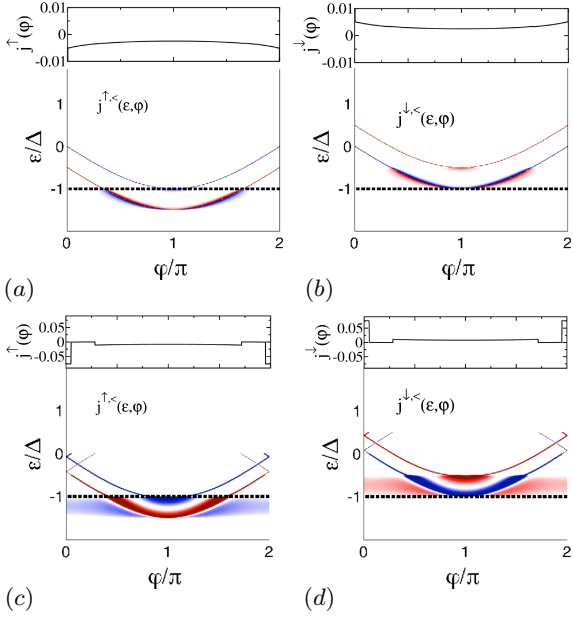
\begin{figure}
  \begin{tabular}{lr lr}
 \scalebox{0.07}{\includegraphics{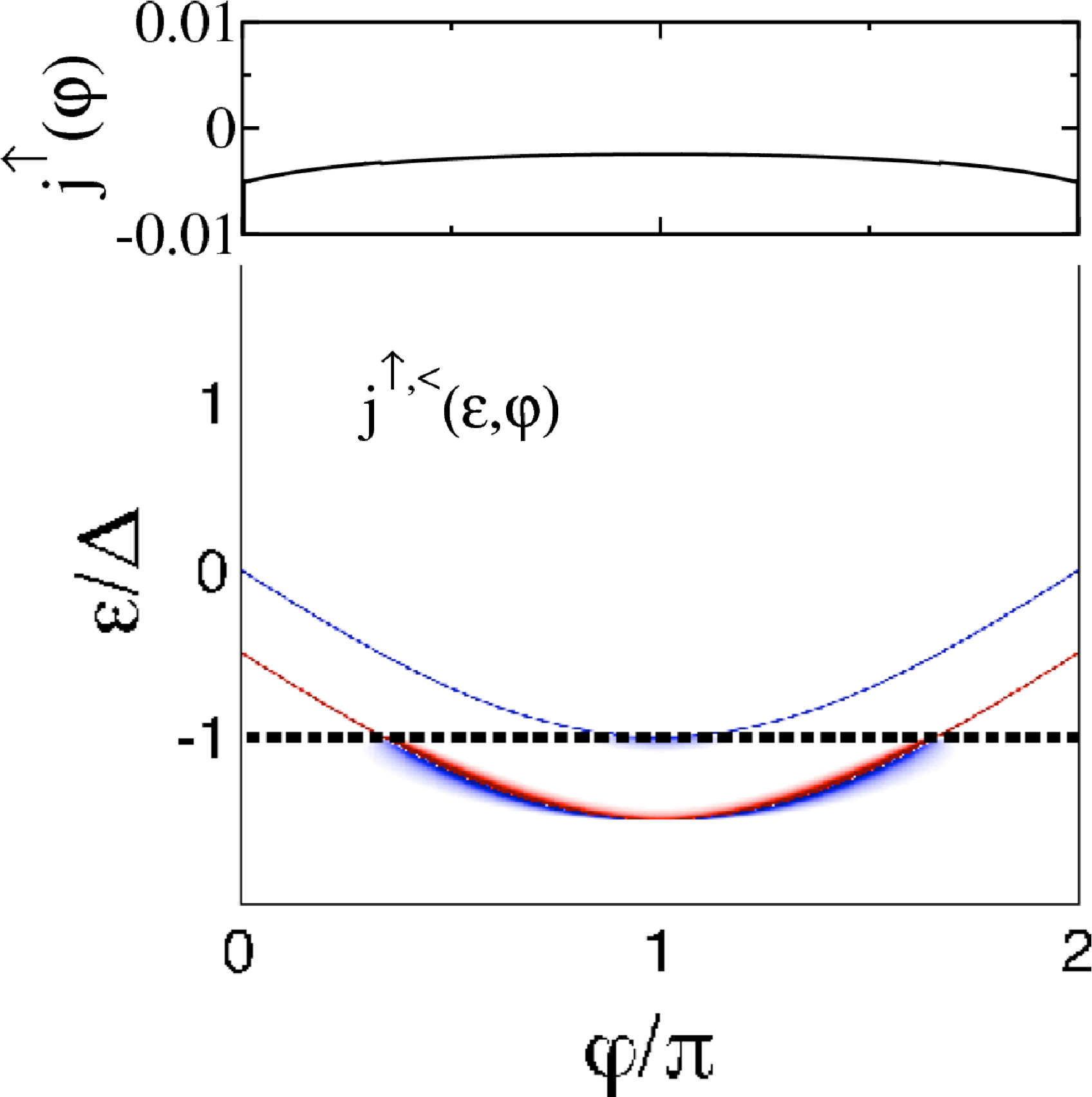}}  & \scalebox{0.07}{\includegraphics{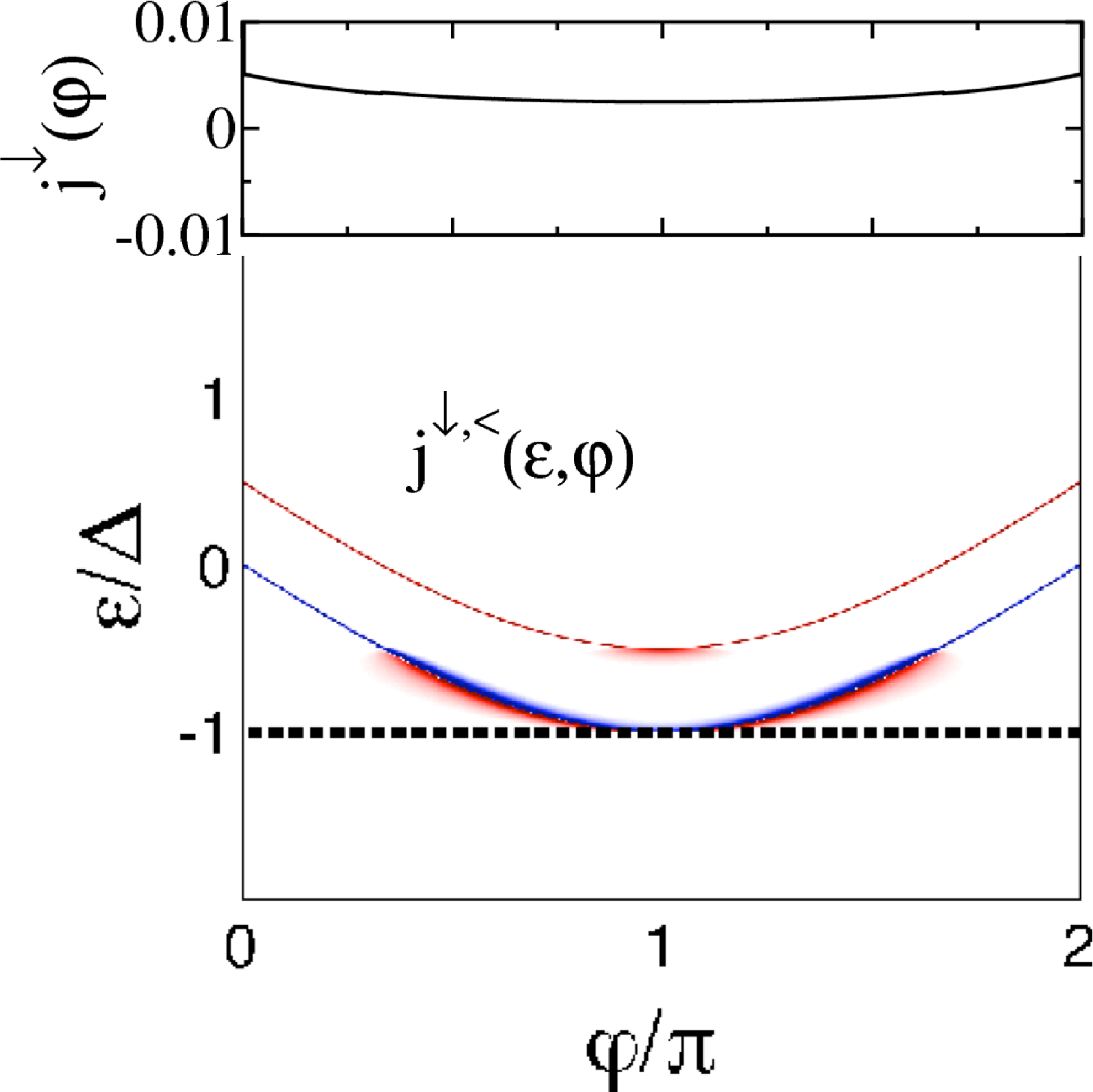}}   \\
 \begin{picture}(100,0)
\put(5,5){\makebox(0,3){$(a)$}}
\end{picture} & \begin{picture}(100,0)
\put(5,5){\makebox(0,3){$(b)$}}
\end{picture} \\
\scalebox{0.07}{\includegraphics{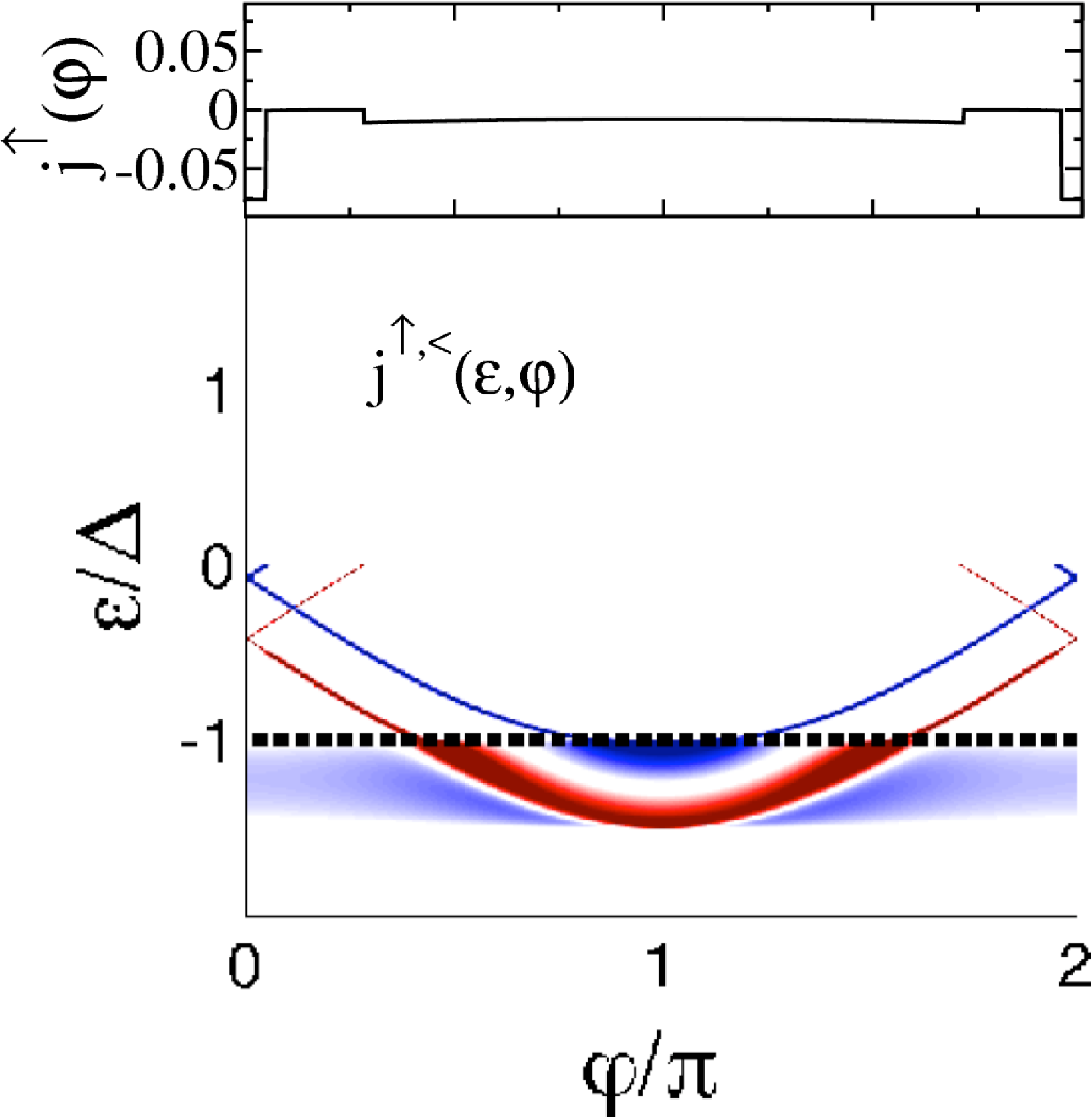}}  & \scalebox{0.07}{\includegraphics{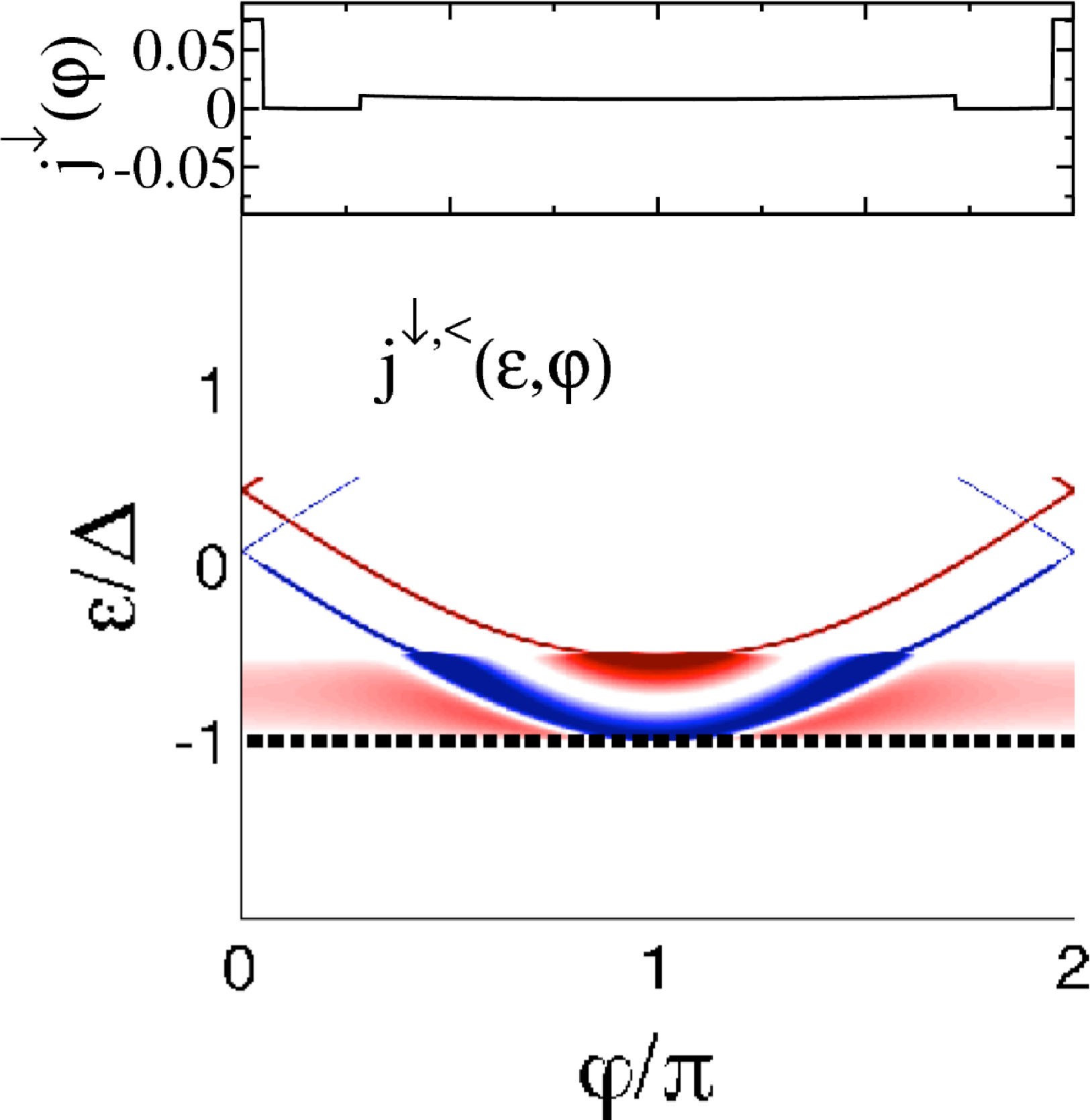}}   \\
 \begin{picture}(100,0)
\put(5,5){\makebox(0,3){$(c)$}}
\end{picture} & \begin{picture}(100,0)
\put(5,5){\makebox(0,3){$(d)$}}
\end{picture} \\
  \end{tabular}
\caption{\label{small theta spin current}
(Color online) The spin current kernel, (a) $j^{\uparrow,<}$ and (b) $j^{\downarrow,<}$, for a $\pi$ junction ($v_0=0, {\cal D}_S=1$) with tilt angle $\vartheta=0.1\pi/2$. Panels (c) and (d) show $j^{\uparrow,<}$ and $j^{\downarrow,<}$, respectively, for tilt angle $\vartheta=\pi/4$. The precessing spin has a frequency of $\omega_L=0.5\Delta$ and the temperature $T=0$. The resulting spin-current-phase relation is plotted in units of $\Delta$ on top of each panel. As seen the phase-dependent population of the scattering states makes the spin-current an almost steplike function of phase as was the case for the charge-current-phase relation shown in figure \ref{spectralcurrentplots}.}
\end{figure}

Contrary to the charge current case, there are first-order contributions to the spin current if the tilt angle is assumed to be small, i.e. $\sin \vartheta\approx\vartheta$. In section \ref{section:small theta charge current}, it was found that for a small angle $\vartheta$, the change in the t matrix $\delta t$ compared to the t matrix for zero tilt angle, $\check{t}^0_\alpha$, i.e. $\check t_\alpha=\check t^0_\alpha+\delta \check t_\alpha$, was given by $\delta \check t_\alpha={\rm e}^{-i\omega_L t} \delta \check t^\uparrow_\alpha+{\rm e}^{i\omega_L t} \delta \check t^\downarrow_\alpha$. Hence, the triplet t matrix is $\check t^t_\alpha=\delta \check t_\alpha$ giving a spin current matrix of
\begin{equation}
\hat{j}^{t,<}_\alpha=2\pi i[\delta \check t^t_\alpha,\check g^0_\alpha]^<_\circ.
\end{equation}
The z component of the spin current is zero since $\delta \check t^d_\alpha=0$ (as was found in section \ref{section:small theta charge current}). Instead, the spin current consists of the two components
\begin{equation}
\jsvec^{s}_\alpha(t)={\rm e}^{-i\omega_L t}\delta \jsvec^{\uparrow}_\alpha + {\rm e}^{i\omega_L t}\delta \jsvec^\downarrow_\alpha
\end{equation}
scaling linearly with $\vartheta\ll\pi/2$ and where 
\begin{equation}
\delta \jsvec^{\uparrow/\downarrow}_\alpha=\frac{1}{4} \int \frac{d \varepsilon}{8 \pi i}2\pi i \mbox{Tr} [ \hat{\tau}_3 \hat{\svec} [\delta \check{t}^{\uparrow/\downarrow}_\alpha(\varepsilon,t),\check{g}^0_\alpha(\varepsilon)]^<_\circ ].
\end{equation}
The upper two panels in figure \ref{small theta spin current} show the current kernels $j^{\uparrow,<} (\varepsilon, \varphi)$ and $j^{\downarrow,<} (\varepsilon, \varphi)$ and the resulting spin-current-phase relation for a small tilt angle, $\vartheta=0.1\pi /2$. The two scattering states carry spin currents in opposite directions as opposed to the charge current Andreev levels which carry a charge in the same direction, see figure \ref{spectralcurrentplots} (a).

\subsubsection{Arbitrary tilt angle}

Increasing the tilt angle modifies only the magnitude of the spin current. The scattering into the sidebands increases as seen in figure \ref{small theta spin current} (c) and (d) while the z component of the spin current remains zero. Consequently, the spin polarization still rotates with the precession frequency $\omega_L$ and can be written as $\jsvec_\alpha(t)=\jsvec_\alpha^\uparrow {\rm e}^{-i \omega_Lt}+\jsvec_\alpha^\downarrow {\rm e}^{i \omega_Lt}$ where $\jsvec^{\uparrow/\downarrow}_{\alpha}$ takes the form
\begin{equation}\label{j up-down commutator integral}
\jsvec^{\uparrow/\downarrow}_{\alpha}(t)=\frac{1}{4}\int \frac{d \varepsilon}{8 \pi i}2\pi i \mbox{Tr} [ \hat{\tau}_3 \hat{\svec} [\check{t}^{\uparrow/\downarrow}_\alpha(\varepsilon,t),\check{g}^0_\alpha(\varepsilon)]^<_\circ ].
\end{equation}

Similarly to the charge carrying density of states (figure \ref{spectralcurrentplots}), the scattering into the continuum states increases with increasing tilt angle. Figure \ref{fig:spin currents (up,down,z)} shows the integrated spin currents, i.e. the current-phase relations for the $\uparrow$ and $\downarrow$ components of the spin currents for a transparent junction with $v_0=0$ and ${\cal D}_S=1.0$. The abrupt jumps in the spin-current-phase relations result from the loss or gain of population of spin scattering states as the phase difference $\varphi$ changes shown in figure \ref{small theta spin current}.

\begin{figure}[t]
\includegraphics[width=0.99\columnwidth,angle=0]{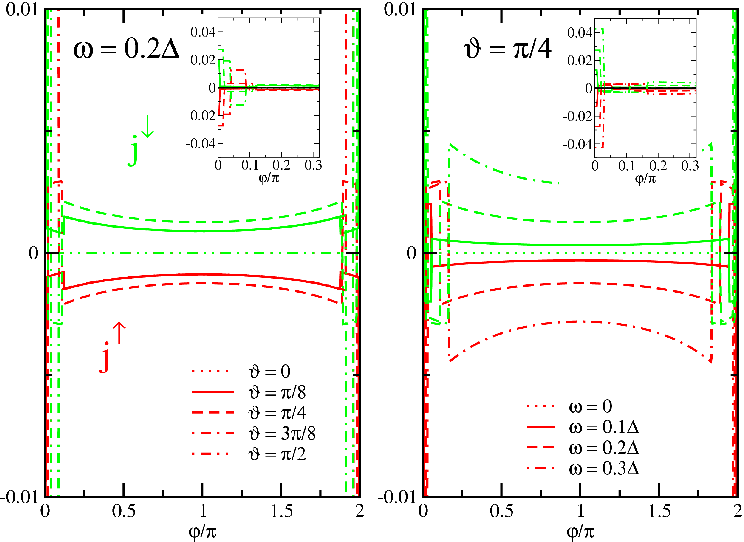}
\begin{picture}(220,0)
\put(10,5){\makebox(0,3){$(a)$}}
\put(130,5){\makebox(0,3){$(b)$}}
\end{picture}\caption{
(Color online) Current-phase relations for the spin-current components $(j^\uparrow,j^\downarrow)$ for $v_0=0$ and ${\cal D}_S=1.0$. 
The time-independent $j^z$-component of the spin current is zero for all cases studied (full black line). The precession frequency is $\omega=0.2 \Delta$, the tilt angle $\vartheta=\pi/4$ and the temperature $T=0$. The spin currents are plotted in units of $\Delta$.}
\label{fig:spin currents (up,down,z)}
\vspace*{-0.3truecm}
\end{figure}
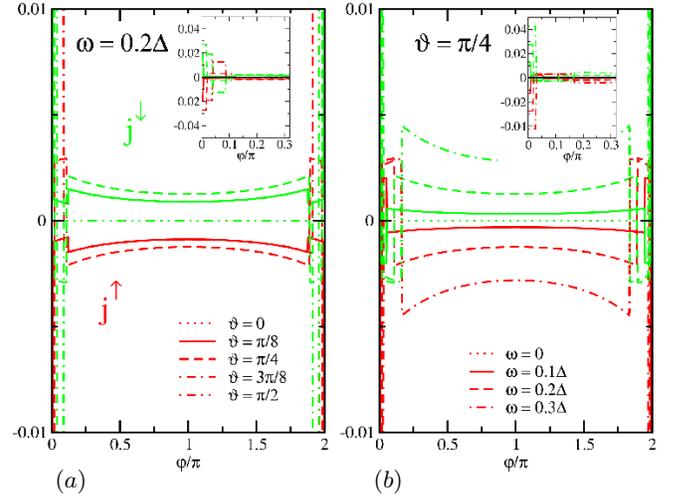

Taking a closer look at the triplet commutator and using equation (\ref{t triplet}), the spin-current matrix is (suppressing the index $\alpha$ for the leads)
\begin{eqnarray}\label{spin commutator}
\hat{j}^{t,<}&=&2\pi i[\check{t}^t,\check{g}^0]^<.
\end{eqnarray}
Dividing the retarded t matrix into an anomalous t matrix,
\begin{equation}\label{t A split}
\hat{\phi}^{t,R}_A=\left(\begin{array}{cc} 0 & (\phi^{t,R}_A) i\sy \\ i\sy ({\tilde 
\phi}^{t,R}_A) & 0  \end{array}\right),
\end{equation}
and "normal" t matrix,
\begin{equation}\label{t N split}
\hat{\gamma}^{t,R}_N=\left(\begin{array}{cc} \gamma^{t,R}_N & 0 \\ 0 & - \sigma_y ({\tilde \gamma}^{t,R}_N )\sigma_y \end{array}\right),
\end{equation}
(similarly for the Keldysh and advanced components), and doing the same for the lead Green's function $\check{g}^0$
\begin{eqnarray}\label{g NA split}
\hat{f}^{0,R}_A&=&\left(\begin{array}{cc} 0 & (f^{0}_A) i\sy \\ i\sy ({\tilde 
f}^{0}_A) & 0  \end{array}\right) \\ \nonumber
\hat{g}^{0}_N&=&\left(\begin{array}{cc} g^{0}_N & 0 \\ 0 & - \sigma_y ({\tilde g}^{0}_N)\sigma_y \end{array}\right),
\end{eqnarray}
the spin-current matrix is given by
\begin{eqnarray}\label{spin NA commutators}
\hat{j}^{t,<}&=&2\pi i[\check{\gamma}^t_N,\check{g}^0_N]^<_\circ+2\pi i[\check{\phi}^t_A,\check{f}^0_A]^<_\circ \\ \nonumber
&+&2\pi i[\check{\gamma}^t_N,\check{f}^0_A]^<_\circ+2\pi i[\check{\phi}^t_A,\check{g}^0_N]^<_\circ
\end{eqnarray} 
where the first term is a contribution from the normal Green's functions and the second term is given by the anomalous Green's functions. The last two terms do not contribute to the current since their nonzero elements are off-diagonal in Nambu-spin space.
We will separate the spin-current into two contributions, one from the normal parts of Green's function and t matrix ($\jvec^s_N$)  and one from the anomalous parts ($\jvec^s_A$). Thus we write  
\begin{eqnarray}\label{spin currents normal anomalous}
\jvec_N^s&=&\frac{1}{4}  \int \frac{d \varepsilon}{8 \pi i} 2\pi i \mbox{Tr} [ \hat{\tau}_3 \hat{\svec} 
 [\check{\gamma}^t_N(\varepsilon,t),\check{g}^0_N(\varepsilon)]^<_\circ ] \\ \nonumber
 &{\rm and}& \\ \nonumber
\jvec_A^s&=&\frac{1}{4}  \int \frac{d \varepsilon}{8 \pi i} 2\pi i \mbox{Tr} [ \hat{\tau}_3 \hat{\svec} 
 [\check{\phi}^t_A(\varepsilon,t),\check{f}^0_A(\varepsilon)]^<_\circ ].
\end{eqnarray}
The spin current contains contributions from both as can be seen in figure \ref{fig:spin current, NAs} where $\jvec^s_N$ and $\jvec^s_A$ have been plotted as functions of the phase difference in panels (a) and (b). In panel (c), the normal and anomalous contributions to the charge current have been plotted for comparison. As the figure shows, the charge current is completely given by the corresponding anomalous contributions.

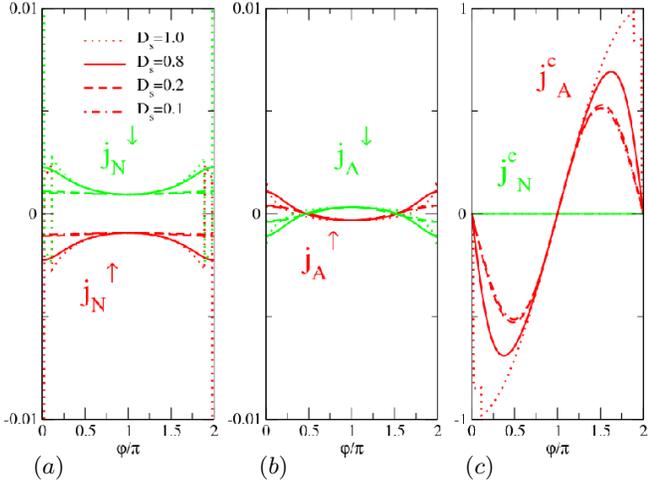
\begin{figure}[t]
\includegraphics[width=0.99\columnwidth,angle=0]{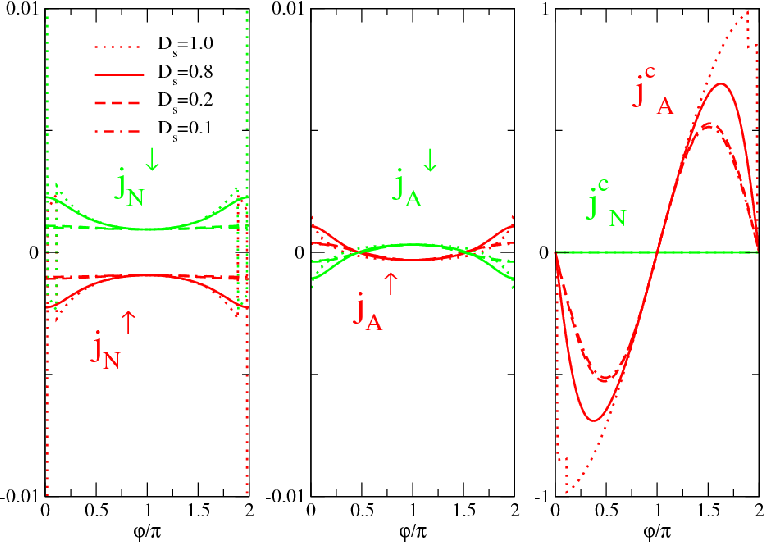}
\begin{picture}(220,0)
\put(5,5){\makebox(0,3){$(a)$}}
\put(90,5){\makebox(0,3){$(b)$}}
\put(168,5){\makebox(0,3){$(c)$}}
\end{picture}
\caption{
(Color online) Spin-current-phase relations for (a) the normal spin current, $j_N$, (b) the anomalous spin current, $j_A$, and (c) the normal and anomalous charge currents, $j_N^c$ and $j_A^c$, for $v_0=0$, precession frequency $\omega=0.2\Delta$, tilt angle $\vartheta=\pi/4$ and temperature $T=0$. The spin currents are plotted in units of $\Delta$ and the charge currents in units of $e\Delta/\hbar$. All currents are scaled by ${\cal D}_S$.}
\label{fig:spin current, NAs}
\vspace*{-0.3truecm}
\end{figure}

The spin current can be expressed in terms of the direction of the rotating spin, $\SMM$, as
\begin{eqnarray}
\jsvec^s(t)&=&\frac{1}{S}{\cal D}_S\beta_H \cos\vartheta\, (\gamma\Heff)\times\SMM(t) \\ \nonumber
&-&\frac{1}{S}\sqrt{{\cal D}_0{\cal D}_S} \beta_\perp \SMM_\perp(t),
\end{eqnarray}
where ${\cal D}_0$ is defined as in section \ref{sec: static spin}.
The components $\beta_H$ and $\beta_\perp$ are plotted in figure \ref{fig:spin current, H, perp, z}. $\beta_\perp$ is finite only when
we have a mixed scattering, i.e. for $v_0>0$. Moreover, $\beta_\perp$ depends quadratically on the precession frequency and has a $\sin\varphi$-like dependence on phase and is zero for both 0 and $\pi$ junctions when in their zero-current or ground state. For high hopping-dependent strengths, ${\cal D}_S=1$ and $v_0=0$, the $\beta_H$ component has a current-phase relation with abrupt jumps which are smoothed out for ${\cal D}_S<1$ and $v_0>0$. In contrast to $\beta_\perp$, the component $\beta_H$ depends linearly on $\omega_L$ for small to intermediate precession frequencies, $\omega_L \lesssim \Delta/2$.

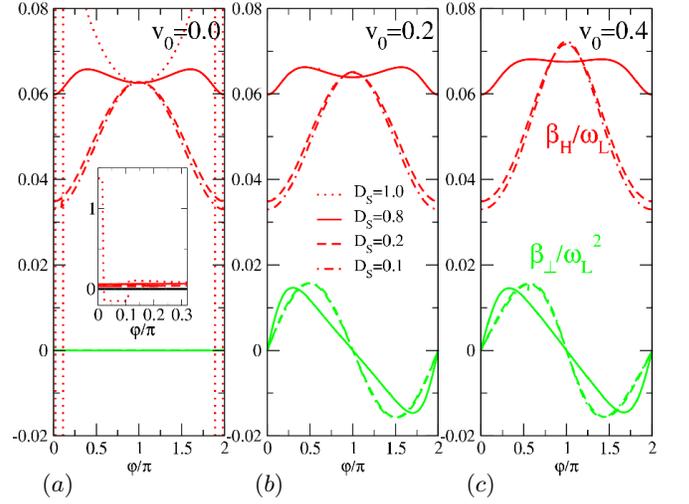
\begin{figure}[t]
\includegraphics[width=0.99\columnwidth,angle=0]{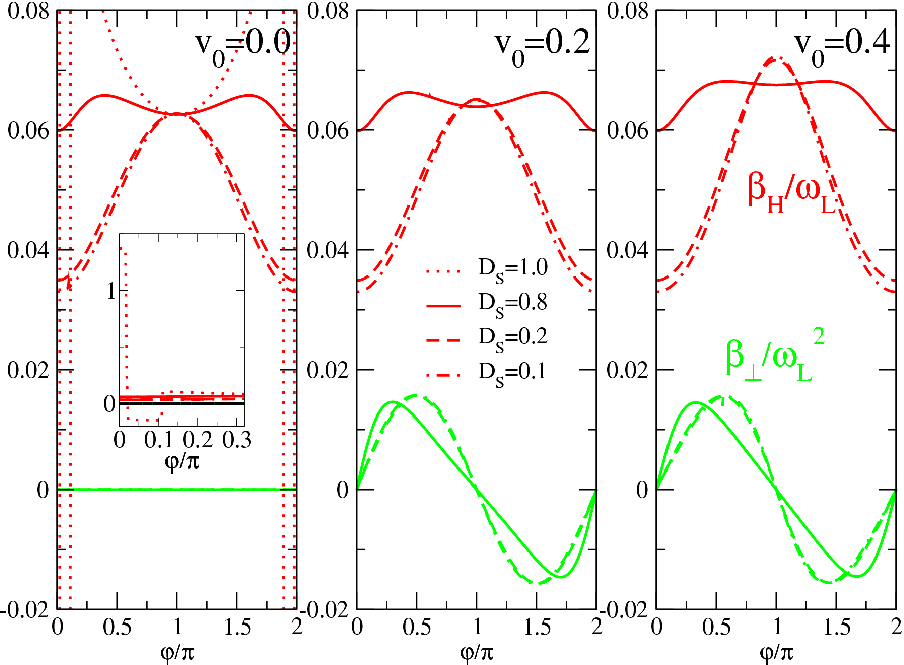}
\begin{picture}(220,0)
\put(5,5){\makebox(0,3){$(a)$}}
\put(85,5){\makebox(0,3){$(b)$}}
\put(165,5){\makebox(0,3){$(c)$}}
\end{picture}
\caption{
(Color online) The phase relation of the functions $\beta_H$ and $\beta_\perp$ at zero temperature for various realizations
of the tunneling strengths. The precession frequency is $\omega=0.2 \Delta$ and the tilt angle is $\vartheta=\pi/4$.
The part of spin current that rotates with the spin precession
$\sim \beta_H$ is always non-zero with a phase-dependent modulation around a mean amplitude. Note the different frequency scalings of 
$\beta_H (\sim \omega_L)$ and $\beta_\perp(\sim \omega_L^2)$.}
\label{fig:spin current, H, perp, z}
\vspace*{-0.3truecm}
\end{figure}

\subsubsection{Tunnel limit}

In the tunnel limit, the triplet t matrix is given by 
$\check t_\alpha^t={\rm e}^{-i\omega_Lt}\check \Gamma^\uparrow_\alpha + {\rm e}^{i\omega_Lt}\check \Gamma^\downarrow_\alpha +\check \Gamma^z_\alpha$. The normal part of the spin current is given by the normal parts of the t matrix as in equation (\ref{spin currents normal anomalous}), such that $\jsvec^s_N={\rm e}^{-i\omega_Lt}\jsvec^\uparrow_N+{\rm e}^{-i\omega_Lt}\jsvec^\downarrow_N$. Performing the integral in equation (\ref{j up-down commutator integral}) assuming that $\omega_L\ll\Delta$ results in
\begin{equation}
j^\uparrow_N\approx \frac{1}{4\pi}v_S^2\sin\vartheta\cos\vartheta\bigg\lbrace{ 4\pi i\Delta-8i\Delta E \left(\frac{\sqrt{3}\omega_L}{2\Delta}\right) \bigg\rbrace}
\end{equation}
where $E(x)$ is a complete elliptic integral of the second kind. A Taylor expansion to second order gives $j^\uparrow_N\approx v_S^2\sin\vartheta\cos\vartheta \frac{3 i \omega_L^2}{16\Delta}$. Similar integrations yield $j^\downarrow_N=-j^\uparrow_N$ and $j_z=0$. The normal spin current at zero temperature is then
\begin{equation}
\jsvec^s_N(t)=\frac{1}{S}v_S^2 \frac{3\omega_L}{16\Delta}\cos\vartheta\,(\gamma\Heff)\times\SMM(t).
\end{equation}
Similarly, the anomalous integrals produce 
\begin{eqnarray}
j^{\uparrow/\downarrow}_A&=& \frac{1}{4\pi}\left(-iv_0v_S\sin\vartheta\sin\varphi \mp v_S^2\sin\vartheta\cos\vartheta\cos\varphi \right) \nonumber \\ 
&\times& \bigg\lbrace{ 8i\Delta K\left(\frac{\omega_L}{2\Delta}\right)-4\pi i\Delta \bigg\rbrace}
\end{eqnarray}
where $K(x)$ is a complete elliptic integral of the first kind. In the limit $\omega_L\ll\Delta$, the $\omega_L$ dependence reduces to $8i\Delta K\left(\frac{\omega_L}{2\Delta}\right)-4\pi i\Delta\approx \pi i\Delta (\omega_L/2\Delta)^2$. Also for the anomalous spin current, $j_z=0$ and
the total anomalous spin current is
\begin{eqnarray}
\jsvec^s_A(t)&=&-\frac{1}{S}v_S^2 \frac{\omega_L\cos\varphi}{16\Delta}\cos\vartheta\,(\gamma\Heff)\times\SMM(t) \\
&-& \frac{1}{S} v_0v_S \frac{\omega_L^2\sin\varphi}{16\Delta} \SMM_\perp(t).
\end{eqnarray}
To summarize, the total spin current is
\begin{equation}
\jsvec^s(t)=\frac{1}{S}{\cal D}_S \beta_H \cos\vartheta\,(\gamma\Heff)\times\SMM(t) - \frac{1}{S} \sqrt{{\cal D}_0{\cal D}_S} \beta_\perp  \SMM_\perp(t)
\end{equation}
where $\beta_H=\frac{1}{64\Delta}[3-\cos\varphi]\omega_L$, $\beta_\perp=\frac{\sin\varphi}{64\Delta} \omega_L^2$ and the transmission coefficients reduce to ${\cal D}_S\approx4v_S^2$ and ${\cal D}_0\approx4v_0^2$ for $v_0, v_S\ll 1$.

\subsubsection{Temperature dependence}

The Josephson spin current described above is an effect of quasiparticles interfering constructively along closed loops when the quasiparticles are subjected to Andreev reflection and transmission across the junction as specified by the rotating classical spin. 
The singlet-pairing nature of the leads does not support spin currents. This means that the spin-dynamics of the Andreev levels will be 
restricted to a small volume (or area) with radius given by the coherence length in vicinity of the nanomagnet junction.
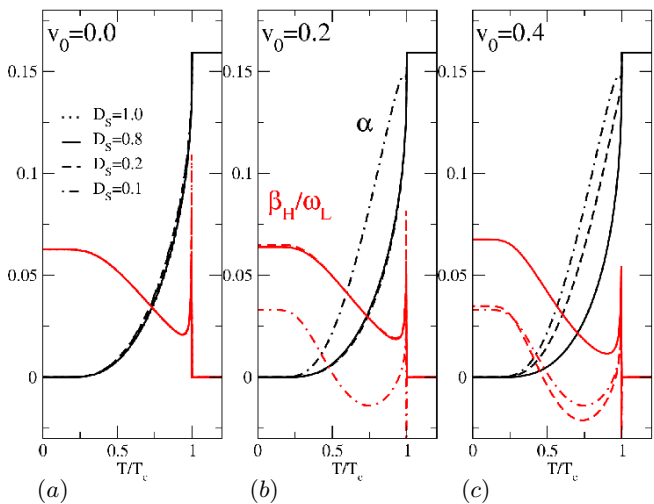
\begin{figure}[t]
\includegraphics[width=0.99\columnwidth,angle=0]{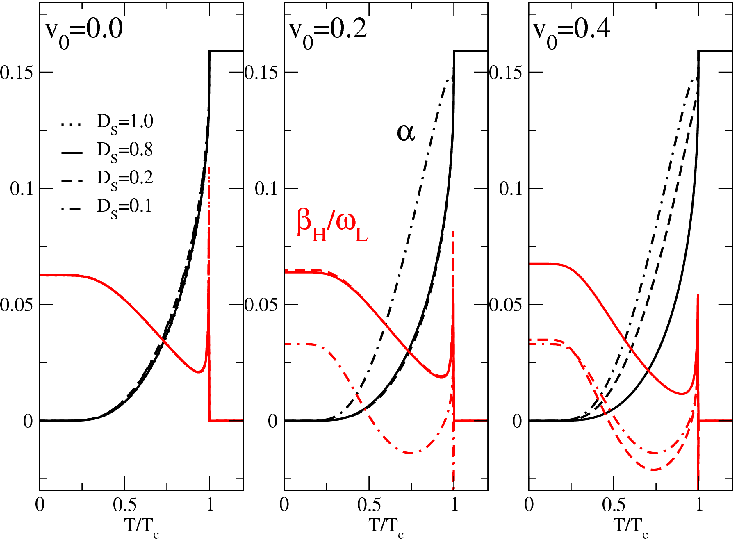}
\begin{picture}(220,0)
\put(5,5){\makebox(0,3){$(a)$}}
\put(85,5){\makebox(0,3){$(b)$}}
\put(165,5){\makebox(0,3){$(c)$}}
\end{picture}
\caption{ 
(Color online) Temperature dependence of the functions $\alpha$, $\beta_H$ and $\beta_\perp$ for precession frequency $\omega_L=0.2\Delta_{T=0}$ and tilt angle $\vartheta=\pi/4$. The functions are plotted with phase difference $\varphi=\pi$ or $\varphi=0$ depending on if the junction is in a $\pi$ state or a $0$ state. Note the strong response at low temperatures captured in $\beta_H$.
At these temperatures the quasiparticle-caused Gilbert damping is frozen out, $\alpha(T\rightarrow0)=0$, and the Andreev-level dynamics
alone will couple back to the precessing spin.
The abrupt jumps at $T\lesssim T_c$ are when $\omega_L=\Delta(T)/2$ and the spin scattering connects the square-root singularities in the superconducting density of states at $\varepsilon=\Delta(T)$.}
\label{fig: IT_lPerpZ}
\vspace*{-0.3truecm}
\end{figure}

However, there is another source of spin current not dependent on the Andreev scattering proceses.
At temperatures $T>0$, the normal contribution to the spin current, $\jsvec^s_N$ given by equation (\ref{spin currents normal anomalous}), results in an extra term
\begin{equation}\label{spin current normal}
\jsvec^s_{qp} = \frac{1}{S^2}{\cal D}_S \alpha \dot{\SMM}(t)\times\SMM(t)
\end{equation}
which is a spin current carried by quasiparticles and is therefore independent of the superconducting phase difference $\varphi$. The parameter $\alpha$ depends in general on the precession frequency $\omega_L$ and the temperature $T$. This contribution to the spin current, $\jsvec^s_{qp}$, may have a finite spin-polarized dc component along the z axis. This situation is similar to a ferromagnetic quantum dot or layer coupled to normally conducting leads where a precession of the magnetization leads to the junction behaving as a spin pump \cite{wang2003,tserkovnyak2002}. In the tunnel limit, it can be shown that the parameter $\alpha=1/2\pi$.

In figure \ref{fig: IT_lPerpZ}, the temperature dependence of the function $\alpha(T,\omega_L)$ related to quasiparticles is plotted as well as the function $\beta_H(T,\omega_L)$ which is related to the Andreev-level dynamics. At temperatures above the critical temperature, $T>T_c$, the spin current is completely given by the quasiparticle spin current $\jsvec^s_{qp}$ and $\alpha=1/2\pi$. As the temperature decreases, the normal quasiparticles freeze out as the superconducting gap opens and $\alpha\rightarrow 0$ as $T/T_c\rightarrow 0$. The functions $\alpha(T,\omega_L)$ and $\beta_{H}(T,\omega_L)$ depend on the hopping amplitudes $v_S,v_0$ as well as on the junction state, i.e. the superconducting phase difference. For a $\pi$ junction, whose tunneling is dominated by $v_S$, $\alpha(T,\omega_L)$ and $\beta_H(T,\omega_L)$ are only weakly dependent on $v_S$ and $v_0$. The function $\beta_\perp(T,\omega_L)$ has a sinusoidal $\varphi$ dependence and since it is zero for both $\pi$ and $0$ junctions it is not shown in figure \ref{fig: IT_lPerpZ}. The parameter $\beta_H(T,\omega_L)$ is zero above the critical temperature and increases as $T/T_c\rightarrow 0$, saturating at $\sim \frac{\pi}{8}\omega_L \alpha_{T>T_c}$. For $0$ junctions, where $v_0<v_S$, the reduction of $\alpha(T,\omega_L)$ is slower. At the same time, $\beta_H(T,\omega_L)$ saturates at a lower value, $\sim\frac{\pi}{16}\omega_L\alpha_{T>T_c}$. The parameter $\alpha(T,\omega_L)$ does not depend on the precession frequency for $\omega_L\lesssim\Delta/2$.

\subsection{Triplet correlations}
\label{sec: triplet correlations}

In section \ref{sec: spin currents}, it was shown that a rotating classical spin inside a phase-biased Josephson junction produces a time-dependent spin current. This is somewhat surprising since the Josephson junction was assumed to consist of two superconducting leads with s-wave symmetry. 
The Andreev processes depicted in figure \ref{Scattering processes} (b) produced by the rotating spin lead to new spin-pairing correlations which are formed when these scattering processes result in positive interference along closed loops. The additional pairing correlations to the usual spin-singlet ones 
$\sim \frac{1}{2}\langle\psi_\uparrow\psi_\downarrow-\psi_\downarrow\psi_\uparrow\rangle$ are the spin-triplet components $\frac{1}{2}\langle\psi_\uparrow\psi_\downarrow+\psi_\downarrow\psi_\uparrow\rangle$, $\langle\psi_\uparrow\psi_\uparrow\rangle$ and $\langle\psi_\downarrow\psi_\downarrow\rangle$. The induced triplet correlations can be expected to form near the junction due to the spin mixing and locally broken spin-rotation symmetry provided 
by $\SMM (t)$ \cite{eschrig2008,houzet08}. A similar situation exists in SFS junctions with conical ferromagnets; spin currents arise due to spin-triplet correlations induced by a helical rotation of the magnetization direction in the ferromagnetic layer \cite{alidoust2010}.
The spin-triplet correlations induced by the rotating spin are localized near the junction and are evanescent on 
length scales on the order of the superconducting coherence length, $\xi_0=\hbar v_f/2 \pi T_c$. The formation of 
triplet Cooper pairs depends on the details of the scattering off and the tunneling over the precessing spin, such as  the precession frequency, $\omega_L$,
the tilt angle, $\vartheta$, and the relative amplitude of hopping strenghts, $v_0,v_S$. It also depends on the leads through the superconducting phase difference $\varphi$, and the temperature, $T$.
\begin{figure*}[t]
\includegraphics[width=1.90\columnwidth,angle=0]{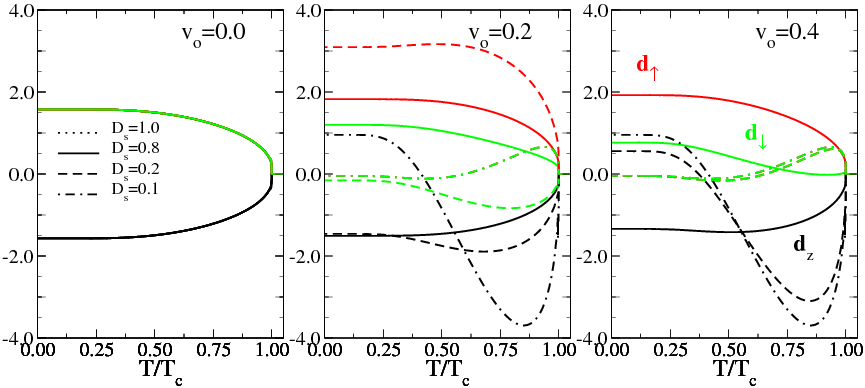}
\begin{picture}(220,0)
\end{picture}
\caption{ 
The components of the $\dvec$ vector, $(\dvec_{\uparrow},\dvec_{\downarrow},\dvec_z)$, as functions of temperature. 
The angle $\vartheta$ between $\Heff$ and $\SMM$ is $\frac{\pi}{4}$ and the precession frequency is 
$0.2\Delta_{T=0}$. In each panel, the spin-independent hopping amplitude $v_o$ is fixed while the spin-transmission 
coefficient ${\cal{D}}_{s}$ is varied. The components are scaled by ${\cal{D}}_{s}\omega_L$. In the left panel all junctions are $\pi$-junctions
while in the center and right panel junctions with ${\cal{D}}_{s}=0.1$ and ${\cal{D}}_{s}=0.1,0.2$ respectively
are 0-junctions. For the 0-junctions the corresponding $\dvec$-vector measures
{\em even-in $\hat{k}$ odd-in $\varepsilon$} spin-triplet correlations}
\label{fig:d-vectors}
\vspace*{-0.3truecm}
\end{figure*}

We now want to quantify the pairing correlations generated in vicinity of the precessing spin. Turning first to the spin-singlet components
we extract the anomalous Green's functions are $f^<_s(\pm k,\varepsilon)$ from the matrix 
$[\check{g}^{in,s}+\check{g}^{out,s}]^<=2 [\check{g}^0_L \circ \left( \check 1 + [\check{t}^s_L,\check{g}^0_L]_\circ \right)]^<$ and write
\begin{equation}
\psi(\hat{k})=\int_{-\varepsilon_c}^{\varepsilon_c} \frac{d \varepsilon}{8 \pi i} \lbrack f^<_s(\hat k,\varepsilon)+f^<_s(-\hat k,\varepsilon) \rbrack
\end{equation}
with "$+(-)$" referring to the anomalous scalar matrix component of the incoming (outgoing) propagator on the left side of the junction and vice versa for the right side. $\psi(\hat{k})$ is a measure of the (singlet) paring correlations available to form a singlet order parameter as 
$\Delta_s(\hat{k})= \lambda_s \eta(\hat{k}) \langle  \eta(\hat{k}^\prime) \psi(\hat{k}^\prime)\rangle_{\hat{k}^\prime\cdot \hat{n}>0}$, where
$\eta(\hat{k})=\eta(-\hat{k})$ are basis functions of even parity on which the pairing interaction may be expanded. 
The energy $\varepsilon_c$ is the usual cut-off that appears in the BCS gap equation (\ref{BCSgapequation}).

The anomalous component of the surface propagator also has spin-vector parts, $\fvec_t^{in}$ and $\fvec_t^{out}$, resulting from the spin-scattering processes in the junction. These induced anomalous components can be quantified in terms of a $\dvec$ vector, which is customary to define in relation to triplet correlations. In general, a $2\times2$ triplet order parameter is given by $\Delta_{\hat{k}}=\dvec(\hat k)\cdot\svec i \sigma_y$ and the $\dvec$ vector points along the direction of zero spin projection of the Cooper pairs \cite{he3book}.
For $\pi$ junctions, we define a vector $\dvec_o$ which is odd in momentum and even in energy as
\begin{equation}
\dvec_{o}(\hat k)= \hat n\!\cdot\! \hat k \int_{-
\varepsilon_c}^{\varepsilon_c} \frac{d \varepsilon}{8 \pi i} 
\lbrack \fvec^<_t(\hat k,\varepsilon)-\fvec^<_t(-\hat k,\varepsilon) \rbrack
\label{odd triplet correlations}
\end{equation}
where the direction of the surface normal is $\hat n$ and, for the left side of the interface, $\fvec^<_t(\pm\hat k,\varepsilon)$ refers to the incoming (outgoing) propagator. For $0$ junctions, we instead define a $\dvec$ vector which is even in momentum and odd in energy, $\dvec_e$, as
\begin{equation}
\dvec_{ e}(\hat k)= \int_{-\varepsilon_c}^{\varepsilon_c} \frac{d \varepsilon}{8 \pi i} s_\varepsilon \lbrack \fvec^K_t(\hat k,\varepsilon)+ \fvec^K_t(-\hat k,\varepsilon) \rbrack
\label{triplet correlations}
\end{equation}
where $s_\varepsilon$ is the sign of the energy $\varepsilon$. This definition is based on the anomalous correlations of $[\check{g}^{in,t}+\check{g}^{out,t}]^<$ which is related to the matrix in equation (\ref{spin commutator 1}) by $[\check{g}^{in,t}+\check{g}^{out,t}]^<=2 [\check{g}^0_L \circ \left( \check 1 + [\check{t}^t_L,\check{g}^0_L]_\circ \right)]^<$.
Spin-triplet pairing that is {\em even-in $\hat{k}$} and {\em odd-in $\varepsilon$} was first considered as a candidate pairing state for $^3$He 
and is in principle not forbidden by symmetry\cite{berezinskii1974} although not realized for superfluid $^3$He.

The triplet correlations span the spin space in 
such a way that $\fvec^K_{z}\sim\frac{1}{2}\langle \psi_\uparrow \psi_\downarrow+ \psi_\downarrow\psi_\uparrow 
\rangle$, $\fvec^K_{\uparrow}\sim \langle \psi_\uparrow \psi_\uparrow \rangle$ and $\fvec^K_{\downarrow}\sim\langle 
\psi_\downarrow \psi_\downarrow \rangle$. Moreover, the instantaneous spin direction of the triplet correlations 
depends on the rotating spin $\SMM$, leading to a time dependence for the $\dvec$ vector of the form 
\begin{equation}
\dvec(t)= \dvec_z + \dvec_\uparrow {\rm e}^{-i \omega_Lt}+\dvec_\downarrow {\rm e}^{i \omega_Lt}.
\end{equation}
The components of the $\dvec$ vector are plotted in figure \ref{fig:d-vectors}. As expected, the components are finite when the leads are 
superconducting ($T/T_c<1$). Setting $v_o=0$ leads to the magnitudes of the components being equal 
except for a scaling of ${\cal{D}}_{s}\omega_L$. These properties are modified for finite $v_o$; as the 
spin-independent tunneling is increased, an asymmetry between $\dvec_{\uparrow}$ and $\dvec_{\downarrow}$ 
emerges and the universal scaling disappears. For low temperatures, $T/T_c\lesssim0.1$, one can express the $\dvec
$ vector in terms of the direction of the rotating spin $\SMM$ as
\begin{equation}
\dvec(t)=\delta_L \dot\SMM(t)\!\times\!\SMM(t)+\delta_H (\gamma \Heff)\!\times\!\SMM(t)+\delta_z \SMM_z.
\label{deltadvec}
\end{equation}
For $\pi$ junctions, the term $\delta_z$ is zero, while $\delta_H=0$ for $0$ junctions. In the tunnel limit, $v_o,v_s\ll1$, 
it is possible to find analytical expressions for these $\SMM$-dependent components. For the odd $\dvec$ vector in 
the tunnel limit at low temperatures, $\delta_{L,o} = \pi{\cal{D}}_{s}\sin(\varphi/2)$ and $\delta_{H,o} = 4\pi iv_ov_s
\sin(\varphi/2)$. 
The even-in momentum $\dvec$ vector, on the other hand, is cut-off dependent and diverges 
logarithmically with $\varepsilon_c$. For the plots in figure \ref{fig:d-vectors}, we used $\varepsilon_c=20\Delta$.
Furthermore, the relation between $\dvec$ vectors on either side of the interface is $\dvec_{R}(t)=-\dvec_L(t)$.

\subsection{Back-action on the precessing spin}

The spin currents on either side of the interface are related by $\jsvec^{s}_{R}(t,\varphi)=-\jsvec^{s}_{L}(t,-\varphi)$, a 
difference leading to a torque, $\torque(t)=\jsvec^{s}_{L}(t)-\jsvec^{s}_{R}(t)$, exerted on the rotating spin $\SMM$. The Josephson spin current consists of two spin-vector components, $\jsvec^s_{L/R}=j_{H,L/R}(\gamma \Heff)\times\SMM + j_{\perp, L/R}\SMM_\perp$. The perpendicular spin currents on either side of the interface, $j_{\perp, L}\SMM_\perp$ and $j_{\perp, R}\SMM_\perp$, are equal and therefore cancel. The other spin currents, $j_{H, L}(\gamma \Heff)\times\SMM$ and $j_{H, R}(\gamma \Heff)\times\SMM$, are equal in magnitude but carry spin angular momentum in opposite directions, leading to a torque, here called Andreev torque, which is given for a single conduction channel by
\begin{eqnarray}
\torque_A(t)=\frac{2\hbar}{S}{\cal{D}}_{s}\beta_H\cos\vartheta \, (\gamma\Heff)\!\times\! \SMM(t).
\label{torque A}
\end{eqnarray}
The torque is parallel to the one generated by the external magnetic field $\Heff$ and hence leads to a 
shift in the precession frequency, $\omega_L\rightarrow \omega_L\lbrack 1+\frac{2\hbar}{S}{\cal{D}}_{s}\beta_H \cos\vartheta\rbrack$. As was seen in section \ref{sec: triplet correlations}, the rotating spin induces local spin-triplet correlations near the junction interface and the spin-triplet correlations allows the superconducting leads to support a spin current even at low temperatures when the quasiparticles
are frozen out. However, the spin current is nothing but transport of spin-angular momentum and the non-conservation of the spin current results in a torque acting on the rotating spin. The shift in precession frequency of the rotating spin is therefore a direct consequence of the induced spin-triplet correlations.

The spin current carried by normal quasiparticles transport angular momentum from the rotating 
spin into the leads resulting in a damping of the spin precession \cite{slonczewski1996,waintal2000}. This process is the main contribution to the Gilbert damping, which has been studied quite extensively (see \onlinecite{tserkovnyak2002} and references therein). As there are many 
possible contributions to the Gilbert damping, it is often entered as a phenomenological parameter \cite{gilbert2004}. Here, the quasiparticle torque, $\torque_{qp}$, is given by the spin current in equation (\ref{spin current normal}) as
\begin{equation}
\torque_{qp}(t)=\frac{2\hbar}{S^2}\alpha{\cal{D}}_{s}  \dot\SMM(t)\times \SMM(t)
\label{normal torque}
\end{equation}
for one conduction channel.
Since the torque is perpendicular to $\SMM(t)$ as well as $\dot\SMM(t)$, 
it leads to an alignment of $\SMM$ with the effective magnetic field $\Heff$. At temperatures above the critical temperature, $\alpha=(2 \pi)^{-1}$,
but as the temperature decreases, the quasiparticles freeze out and the quasiparticle spin current as well as the quasiparticle torque vanish as $T/T_c\rightarrow 0$. This reduction of Gilbert damping due to superconducting phase transitions was investigated in 
references [\onlinecite{bell2008,morten2008}] where the Gilbert damping in domains with a precessing magnetization was measured at 
temperatures around $T_c$.

We propose that the Andreev torque can be measured in the following way:
Assume that a nanoparticle doped with a few magnetic atoms giving it large spin, say $S\sim50\hbar$, is placed between two superconducting aluminum (Al) leads. The superconducting 
coherence length, $\xi_0$ ($\sim2\,\mu$m for Al), limits the area within which the time-dependent spin dynamics affects 
the Andreev reflection to $\sim\pi\xi_0^2$. A contact width $\sim100$ nm results in $n\sim600$ conduction channels. The superconducting gap $\Delta\sim200\,\mu$eV in bulk Al, but can be made considerably smaller in the point contact.
The changes in the power absorption spectrum of an FMR experiment can now be calculated.
The width of the resonance peak constitutes of inhomogeneous broadening, which is due to sample 
imperfections such as anisotropy fields, and homogenous broadening $\Delta H_{\rm hom}$, which is due to Gilbert 
damping. The homogenous broadening may be expressed as \cite{platow1998} $\Delta H_{\rm hom}=\frac{2}{\sqrt{3}} H \alpha_{G}$ where $H=\vert \Heff \vert$ and $\alpha_G=\frac{2\hbar}{S} n \alpha {\cal{D}}_{s}$ is the Gilbert constant 
\cite{gilbert2004}. For a typical magnetic field, $H=60$ mT, the macrospin precession is $\sim1\,\mu$eV corresponding 
to a frequency of $10$ GHz. For a junction with ${\cal{D}}_{s}\sim 0.1$, the freezing out of normal 
quasiparticles decreases the width of the resonance peak and results in a difference in homogeneous broadening on the order of $\Delta H_{\rm hom}(T/T_c>1)-
\Delta H_{\rm hom}(T/T_c\rightarrow 0) \sim 26$ mT. As the temperature is lowered, a shift in the resonance peak $H_0$, corresponding to the shift in precession frequency, appears due to the Andreev torque. The shift can be related to the Gilbert constant as $
\Delta \omega_L / \omega_L=\alpha_G  \beta_H \cos\vartheta$. Choosing $\vartheta=\pi/4$, $\beta_H\sim\frac{1}{16}\frac{\omega_L}{\Delta}$ in the low transparency limit for temperatures $T/T_c<1/4$ which leads to a change in the resonance peak of $\Delta H_0/H_0\sim 2\%$. This ratio could be made higher 
by increasing the ratio $\hbar n/S$ or by increasing 
the junction transparency.

\subsection{Magnetization}

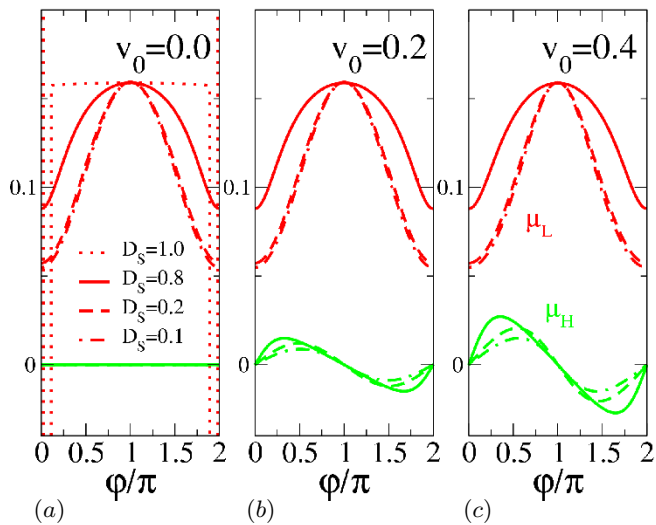
\begin{figure}[t]
\includegraphics[width=0.99\columnwidth,angle=0]{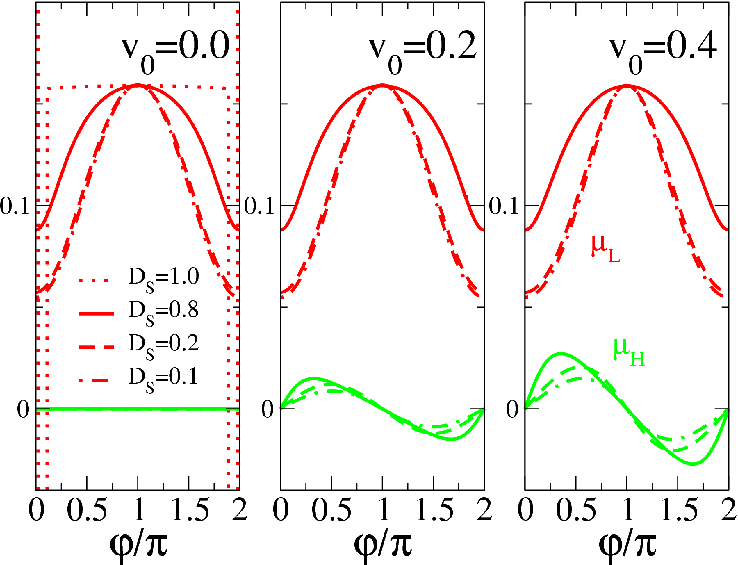}
\begin{picture}(220,0)
\put(3,5){\makebox(0,3){$(a)$}}
\put(83,5){\makebox(0,3){$(b)$}}
\put(165,5){\makebox(0,3){$(c)$}}
\end{picture}
\caption{
(Color online) The phase relation of the functions $\mu_L$ ans $\mu_H$ at zero temperature. The precession frequency is $\omega=0.2 \Delta$ and the tilt angle is $\vartheta=\pi/4$. The function $\mu_H$ is scaled by ${\cal D}_S$.}
\label{fig:magnetization, L, H, z}
\vspace*{-0.3truecm}
\end{figure}

\begin{figure}[t]
\includegraphics[width=0.99\columnwidth,angle=0]{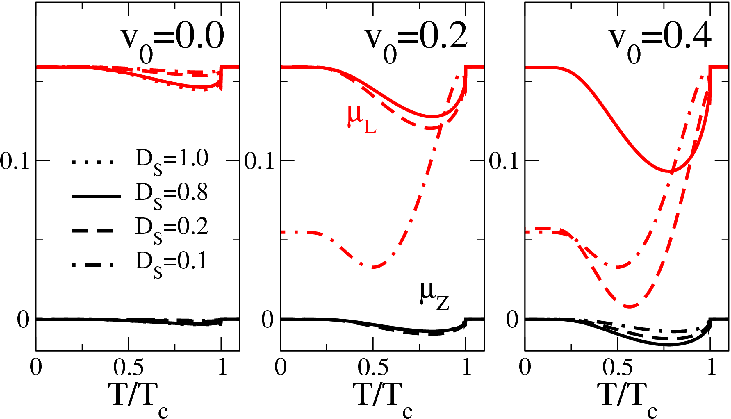}
\begin{picture}(220,0)
\put(3,5){\makebox(0,3){$(a)$}}
\put(83,5){\makebox(0,3){$(b)$}}
\put(165,5){\makebox(0,3){$(c)$}}
\end{picture}
\caption{
(Color online) The temperature dependence of the functions $\mu_L$ and $\mu_z$ at phase differences $\varphi=\pi$ or $\varphi=0$ depending on the state of the junction ($v_S>v_0$ or $v_S<v_0$). The precession frequency and the tilt angle are the same as in figure \ref{fig:magnetization, L, H, z}.}
\label{fig: mag T}
\vspace*{-0.3truecm}
\end{figure}
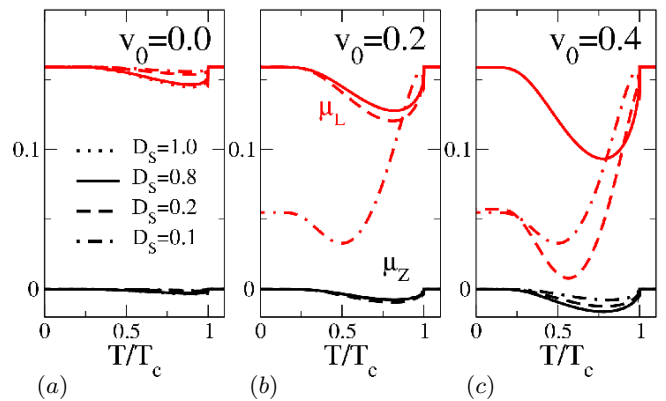

The proximity effect is the penetration of superconducting order into a normal metal \cite{degennes1964} or a ferromagnet \cite{buzdin2004}. The reverse case, the so-called inverse proximity effect, occurs when ferromagnetic order penetrates into the superconductor \cite{bergeret2004,halterman2004}. In the present case, where a precessing classical spin sits between two superconducting leads, there is an induced magnetization in the leads besides the induced triplet correlations. In the case of a static spin, only the spin-singlet correlations and possibly the spin-triplet correlations with spin projection along the direction of the classical spin contribute to the magnetization \cite{bergeret2004}. For a finite precession frequency, also the time-dependent spin-triplet correlations contribute to the magnetization, which can be written on the form
\begin{equation}
\Mvec_\alpha=\Mvec_\alpha^d+\Mvec_\alpha^\uparrow{\rm e}^{-i\omega_L t}+\Mvec_\alpha^\downarrow{\rm e}^{i\omega_L t}.
\end{equation}
The magnetization is given by \cite{serene1983}
\begin{equation}
\Mvec_\alpha(t) =  \frac{\mu_B}{2}  \int \frac{d \varepsilon}{8 \pi i}  \mbox{Tr} [ \hat{\tau}_3 \hat{\svec} 
( \hat{g}^{i,<}_{\alpha}(\varepsilon,t) + \hat{g}^{o,<}_{\alpha}(\varepsilon,t)) ],
\label{mag}
\end{equation}
where $\mu_B$ is the Bohr magneton. From numerical investigations it can be found that
\begin{equation}
\Mag_L(t)=\frac{\mu_B}{S^2} {\cal{D}}_{s}\mu_L \dot{\SMM} \times \SMM + \frac{\mu_B}{S}\mu_H(\gamma \Heff)\times\SMM
\end{equation}
where $\mu_L$, $\mu_H$ and $\mu_Z$ are functions of temperature, phase difference and the hopping amplitudes $v_S$ and $v_0$. The low-temperature phase relation of the parameters $\mu_L$ and $\mu_H$ are plotted in figure \ref{fig:magnetization, L, H, z}. As low temperatures $\mu_Z$ is zero. The parameter $\mu_H$ has a sinusoidal phase relation while $\mu_L$ shows a cosine behavior. In the tunnel limit, $\mu_L=\frac{1}{6\pi}(2-\cos\varphi)$, $\mu_H=\frac{2}{3\pi}v_0v_S\sin\varphi$ while $\mu_Z=0$. The temperature dependence of the parameters $\mu_L$ and $\mu_Z$ for a fixed phase difference is shown in figure \ref{fig: mag T}. The phase difference is $\varphi=\pi$ if the junction is in a $\pi$ state ($v_S>v_0$) or a $0$ state ($v_S<v_0$). For these phase differences, the parameter $\mu_H$ (not shown) is zero. The parameter $\mu_Z$, on the other hand, is finite for temperatures $1/2<T/T_c<1$. The parameter $\mu_L\sim(2\pi)^{-1}$ as $T\rightarrow0$ for $\pi$ junctions while $\mu_L\sim(6\pi)^{-1}$ for $0$ junctions. Note that unlike reference [\onlinecite{bergeret2004}], the magnetization is proportional to the precession frequency and that $\Mvec=0$ for a static spin.

\section{Conclusions}
\label{Sec.Conclusions}

In this work we have presented a detailed analysis of the coupling between the Josephson effect and a nanomagnet. The magnetization of the nanomagnet is brought into precession by an applied external magnetic field. The magnetization of the nanomagnet is modeled as a classical spin and the precession provides a time-dependent as well as spin-active interface which we handled by formulating a boundary condition within the quasiclassical theory of superconductivity. The boundary condition was formulated in terms of a t-matrix equation describing the scattering between the two superconducting leads \cite{cuevas01,kopu04}. We showed that the t-matrix equation could be solved in the laboratory frame by preserving the explicit time dependence or in a rotating frame where the time dependence has been removed by a unitary transformation. The rotating frame solution is more numerically efficient although it introduces a shift of the chemical energies of the spin-up and spin-down bands of the leads making the laboratory frame solution more suitable for investigations of modifications to the superconducting state of the leads due to the time-dependent boundary condition.

As was already described in reference [\onlinecite{teber2010}], the Josephson charge current is time independent despite the magnetization dynamics. Here, we focus on the changes to the superconducting state and we show that this stationary solution nevertheless exhibits a nonequilibrium population of the density of states due to spin scattering between the Andreev levels. The nonequilibrium population of the Andreev levels results in a modified current-phase relation and causes abrupt jumps as Andreev levels become populated or unpopulated. The spin current, on the other hand, lacks time-independent components. Instead, the spin polarization of the current precesses around the direction of the external magnetic field.

We have shown that the spin current is due to an Andreev-level dynamics with induced spin-triplet components. The spin-triplet correlations are due to the breaking of spin-rotation symmetry caused by the precessing classical spin. In more detail, the spin-triplet correlations are a result of the combination of Andreev retroreflection of electron- and hole-like quasiparticles at the superconductor surfaces and the spin scattering processes caused by the precessing spin in which the quasiparticles undergo spin flip while exchanging energy with the classical spin. The triplet correlations are evidence of a dynamical inverse proximity effect produced by the time-dependent spin-active barrier.

The spin currents generate a torque on the classical spin changing its dynamics. This back-action is due to Andreev scattering processes and shifts the precession frequency of the classical spin and can in principle be tuned using the superconducting phase difference across the junction. There is also a source of Gilbert damping at finite temperatures due to the presence of quasiparticles \cite{gilbert2004}. The amount of Gilbert damping can be controlled by adjusting the temperature since the quasiparticles freeze out as $T\rightarrow0$.

We suggest that the effects of the Andreev torque can be measured in an FMR experiment similar to that in reference [\onlinecite{bell2008}]. A measured shift in precession frequency would be a probe of the Andreev-level dynamics. Conversely, by controlling the Andreev-level dynamics and the inverse proximity effect, one may control the magnetization dynamics of a nanomagnet \cite{thirion2003} which would be useful in spintronics applications.

\section*{Acknowledgments}
The authors are grateful to D. Feinberg, M. Houzet, T. L\"ofwander, J. Michelsen and V. Shumeiko for stimulating discussions.

\end{document}